\documentclass[preprint,flushrt]{aastex}

\newcommand{\bc}{\begin{center}}
\newcommand{\ec}{\end{center}}
\newcommand{\bea}{\begin{eqnarray}}
\newcommand{\eea}{\end{eqnarray}}

\def\beq{\begin{equation}}
\def\eeq{\end{equation}}

\def\ben{\begin{enumerate}}
\def\een{\end{enumerate}}
\def\bit{\begin{itemize}}
\def\eit{\end{itemize}}
\def\bi{\bibitem}

\def\mpc{\,{\rm {Mpc}}}
\def\kpc{\,{\rm {kpc}}}

\def\kms{\,{\rm {km\, s^{-1}}}}
\def\msun{M_\odot}
\def\gyr{\rm Gyr}

\def\lam{\lambda}
\def\vcir{V_c}
\def\v200{V_{200}}
\def\vd{V_{d}}
\def\vdmax{V_{2.2}}
\def\sab{S_{a,b}}
\def\s23{S_{2,3}}
\def\rvir{r_{200}}
\def\Rd{r_d}
\def\Mh{M_h}
\def\Md{M_d}

\def\md{m_d}

\def\Ldi{L_{d,I}}

\def\Eh{E}
\def\Jh{J_h}

\def\jd{j_d}

\def\fr{f_R}

\def\calI{{\cal I}}
\def\calK{{\cal K}}
\def\zstar{Z_{\star}}
\def\birth{\,{\rm {SFR/<SFR>}}}

\def\H0{H_0}
\def\om0{\Omega_{m,0}}
\def\omb0{\Omega_{b,0}}

\def\oml0{\Omega_{\rm \Lambda,0}}
\def\rhoc{\rho_{\rm crit}}

\def\pc{{\rm PC}}

\bibpunct[,]{(}{)}{;}{a}{}{,}

\begin{document}

\title{From Correlations of Galaxy Properties to the Physics
of Galaxy Formation:\\A Theoretical Framework}

\author{Alberto Conti}
\affil{Department of Physics and
Astronomy, University of Pittsburgh, Pittsburgh, PA 15260}

\author{Barbara S. Ryden \& David H. Weinberg}
\affil{Department of Astronomy, The Ohio State University, 
Columbus, OH 43210}

\email{Email:\sf conti@phyast.pitt.edu; ryden, dhw@astronomy.ohio-state.edu}

\shorttitle{Interpreting Correlations of Galaxy Properties}
\shortauthors{Conti, Ryden, and Weinberg}
\singlespace

\begin{abstract}
Motivated by forthcoming data from the Sloan Digital Sky Survey, we
present a theoretical framework that can be used to interpret Principal
Component Analysis (PCA) of disk galaxy properties. We use the formalism
introduced by Mo, Mao, \& White to compute the observable properties of
galaxies in a number of model populations, varying assumptions about
which physical parameters determine structural quantities and star
formation histories. We then apply PCA to these model populations.  Our
baseline model assumes that halo mass, spin parameter, and formation
redshift are the  governing input parameters and that star formation is
determined by surface  density through a Schmidt law.  To isolate
physical effects, we consider  simplified models in which one of these
input parameters is held fixed. We also consider extended models that
allow variations in disk mass or angular momentum relative to halo
quantities or that choose a star formation  timescale independent of
surface density.  In all cases, the first principal  component is
primarily a measure of the shape of the spectral energy distribution
(SED), and it is usually driven by  variations in the spin parameter,
which influences star formation through  the disk surface density.  The
second and (in some cases) third principal  components consist mainly of
``scale'' parameters like luminosity, disk radius, and circular
velocity.  However, the detailed division of these scale  parameters,
the disk surface brightness, and the rotation curve slope among the
principal components changes significantly from model to model. Our
calculations yield predictions of principal component structure for the
baseline model of disk galaxy formation, and a physical interpretation
of these predictions. They also show that PCA can test the core
assumptions of that model and reveal the presence of additional physical
parameters that may govern observable galaxy properties.
\end{abstract}

\clearpage

\section{Introduction}
Recent progress in the empirical understanding of galaxy formation
has been driven in large part by evolutionary studies, which can 
trace changes in the galaxy population over lookback time.
The properties of galaxies at the present day, and the correlations
among those properties, offer equally important clues to the
physical processes that govern galaxy formation.  This empirical
approach will be revolutionized in the next few years by the
Sloan Digital Sky Survey (SDSS; \citealt{yor2000}) and other large
imaging and spectroscopic surveys, which provide distances and 
multi-color photometric information for large samples of galaxies.
In statistical analyses of such large data sets, Principal Component
Analysis (PCA) is a powerful tool for characterizing correlations 
among many measurable quantities in terms of a few, independently
varying components.  If we think of each galaxy as a point in the
multi-dimensional space of its observable properties, then PCA yields
a compact but highly informative description of the distribution of
galaxies in that multi-dimensional space.  Applications of PCA to
elliptical galaxies have shown that they occupy a thin ``fundamental
plane'' in the space of luminosity, size, and velocity dispersion
\citep{djo1987,dre1987,guz1993}.  Application to disk galaxies has been
more difficult because of the greater difficulty in controlling selection
biases for lower surface brightness objects \citep{dis1983}; however,
there are well established bivariate correlations between luminosity
and rotation velocity \citep{tul1977} and between luminosity and
surface brightness \citep{dej2000,cro2001,bla2001}.
The SDSS will be an ideal target for PCA of all galaxy types.  The goal
of this paper is to provide a theoretical framework for connecting PCA
of disk galaxy properties to the underlying physics that controls
disk galaxy formation.

The conventional sketch of disk galaxy formation has its roots in the
work of \citet{whi1978} and \citet{fal1980}, updated and substantially
extended by \citet{dal1997} and \citeauthor{mo1998} (\citeyear{mo1998},
hereafter MMW; see also \citealt{mao1998,hea1999,mo1999,van2000}, 2001).
A dark matter halo undergoes gravitational collapse and settles into 
dynamical equilibrium at some formation redshift.  The baryonic
material within this halo (or some fraction of it) dissipates its
energy and settles into a gas disk, preserving its specific angular
momentum, and the disk gravity causes the dark halo to contract
adiabatically \citep{blu1986}.  The mass of the disk is determined
by the halo mass and the cooled baryon fraction, and the size of 
the disk is determined by the halo virial radius and the
halo spin parameter $\lambda$.  The initially gaseous disk is
converted into stars according to an empirical correlation between
surface density and star formation rate \citep{sch1959,ken1998}.

Hydrodynamic simulations 
\citep[e.g.,][]{kat1992,nav1994,ste1994,dom1998,kae1998,wei1998,sommer1999,
nav2000} and ``semi-analytic''
methods based on halo merger trees \citep{whi1991,kau1993,col1994,avi1998,
som1999} provide more sophisticated models of disk galaxy formation,
treating some of the underlying physical processes in greater detail.
However, the MMW formalism appears well suited to our purposes here,
at least at this initial stage, because it allows easy variation of its 
input assumptions and relatively straightforward interpretation of its
governing parameters.  Its principal shortcoming is a highly idealized
description of physical processes that are bound to be more complex
in a fully realistic calculation.  Our present application, though, is 
relatively undemanding in terms of quantitative accuracy, since we
wish to calculate correlations among observable properties but do not
consider the zero-points or full distribution functions of these 
properties.  

The ``baseline model'' of disk galaxy formation that we adopt later
in this paper is based on the conventional picture outlined above,
and specifically on the version of that picture presented by MMW.
However, while MMW set out to calculate and test the predictions of
the cold dark matter (CDM) scenario of galaxy formation, making the
most reasonable assumptions that could be implemented within their
computational framework, we seek instead to understand how departures
from the standard assumptions might reveal themselves in the principal
components of galaxies' observable properties.\footnote{The recent
paper by \citet{she2001} is similar in spirit to our investigation,
though it focuses on a more restricted set of models and observables,
and it analyzes model galaxy populations in terms of ``fundamental plane''
relations rather than principal components.}
We will therefore compare predictions of the baseline model to
those of ``extended'' models that allow additional variations in 
physical input parameters or alter the link between galaxy structure
and star formation.  To clarify the links between governing parameters
and predicted correlations, we also consider simplified models in which
variation of some input parameters is suppressed.  We gain further
insight by examining the correlations of principal components with
the physical inputs like halo mass, spin parameter, and formation redshift,
illustrating the connection between the observables and the 
``hidden variables'' of this theory of galaxy formation.

The next section describes our method of calculating galaxy properties,
concluding with a summary of how we go from the physical input parameters
to the observable quantities that we use for PCA.  Details of the
chemical evolution model are presented in an Appendix.
In \S\ref{sec:pca}
we briefly review the ideas behind PCA.  In \S\ref{sec:results} we present
our results for the correlations and principal component structure
of the simplified models, the baseline model, and the extended models,
with a summary in \S\ref{subsec:summary}.
We conclude with a discussion of future directions.

\section{Calculation of Galaxy Properties}
\label{sec:overview}

Our investigation focuses on the population of relatively isolated
disk galaxies.  The fragility of galactic disks implies that such
galaxies must have had fairly quiescent accretion histories
(e.g., \citealt{tot1992,qui1993}).  We therefore adopt the calculational
methods of MMW, which consider the settling of gas within a single
dark matter halo, rather than a more elaborate method based on
halo merger trees.  We incorporate star formation and spectral and
chemical evolution following (and extending) the techniques of
\citet{hea1999}; similar methods have been employed by 
van den Bosch (2000, 2001).  In real galaxies, the processes of
gas dissipation and star formation may be quite complex,
especially if feedback from young stars and supernovae 
plays a major role in regulating gas flow into and out of the 
forming galaxy.  In our calculations, we summarize the effects
of these complex processes with a small number of parameters,
which characterize, for example, the ratios of disk mass and
disk angular momentum to halo mass and halo angular momentum.
These parameters, together with parameters that describe the
properties of the dark halo itself, define an individual galaxy
and allow us to calculate numerous observable properties as
described below.  In \S\ref{sec:results}, we will consider ten
different ``models of galaxy formation,'' which 
differ from each other in the {\it distributions} adopted 
for these galaxy input parameters.

\subsection{Cosmology}
\label{subsec:cosmology}

For all of our calculations, we adopt a cosmological model with
matter density parameter $\om0=0.3$, a cosmological constant
$\oml0=0.7$, a Hubble constant $h\equiv \H0/100 \kms\mpc^{-1}=0.7$,
a cold dark matter power spectrum with inflationary index $n=1$ and
normalization $\sigma_8=1.13$, and a baryon
density $\omb0 h^2 = 0.0125$ \citep{bur2001}.
The values of cosmological parameters have a limited impact on
our calculations.  The matter density, cosmological constant,
and Hubble constant determine
the age of the universe and the relation between time and redshift, 
which in turn influence the spectral evolution of the galaxies.
We adopt the ratio of baryon density to total matter density,
$f_b\equiv \omb0/\om0$,
as an upper limit on the ratio of galaxy disk mass to halo mass.
The amplitude and shape of the matter power spectrum, together with
the growth history determined by $\om0$ and $\oml0$, determine the
distribution of halo masses and the characteristic formation redshift
and concentration parameter at each mass.  Changes to the cosmological
parameters would shift the zero-points of some of our computed
correlations between galaxy properties; for example, by changing
the ages of stellar populations or the contributions of dark matter
to rotation velocities. However, changes within the range allowed by current
observations would probably not have much impact on the
structure of the principal components.

\subsection{Halo Formation}
\label{subsec:ps}

For each dark matter halo, we need to know the mass, the density profile,
the total angular momentum, and the formation redshift.  In combination
with other parameters described later, these determine the mass, size,
rotation curve, and star formation history of the corresponding disk.

We draw halo masses from the analytic mass function derived by
\citet[][ hereafter PS]{pre1974}.
The comoving number density
of dark halos with mass in the range $\Mh\to \Mh+d\Mh$ at
redshift $z$ is 
\beq
\label{eq:ps}
n_h(\Mh,z)d\Mh =
-{{\overline \rho}_0\over \Mh} \left({2\over \pi}\right)^{1/2} 
{\delta_z \over \sigma^2 (r)}
\left(\partial \sigma \over \partial \Mh \right)
{\rm exp} \left\lbrack -{\delta_z^2 
\over 2 \sigma ^2(r) }\right\rbrack 
{ d \Mh},
\eeq
\noindent
where ${\overline \rho}_0$ is the mean mass density of the universe at
the present time, $\delta_z$ is the critical linear overdensity for collapse
at redshift $z$, $\sigma$ is the mass variance of linear density
fluctuations in top-hat windows of comoving radius $r$, and $\Mh$ is
the average mass inside a sphere of comoving radius $r$. 
In this paper, we concentrate only on the properties of present-day
galaxies, and we therefore draw halo masses from the PS mass
function evaluated at $z=0$.  However, our galaxy evolution code
also computes the properties of galaxy populations at other redshifts.

We assume that $M_h$ represents the total mass (dark and baryonic)
in a sphere of radius $\rvir$, within which the mean density is 200 
times the critical density.  This definition implies a relation between
$M_h$ and the halo circular velocity at $\rvir$,
\beq
\label{eq:massh}
\Mh = \frac{4\pi}{3} 200 \rhoc \rvir^3 = \frac{\v200^3}{10GH(z)} \ .
\eeq
In physical units,
\beq
\label{eq:r200_nfw}
\rvir = 163 \left( \frac{M_h}{10^{12} h^{-1}\msun} \right)^{1/3}
\left(\frac{\om0}{\Omega_m(z)}\right)^{1/2} (1+z)^{-1} h^{-1} \kpc \ ,
\eeq
\noindent
and
\beq
\label{eq:v200_nfw}
\v200 = \biggl({G \Mh \over r_{200}}\biggr)^{1/2} =
200 \biggl({r_{200} \over 200 h^{-1} {\rm kpc}}\biggr) \, 
\biggl({\om0 \over \Omega_m(z)}\biggr)^{1/2} (1+z)^{3/2} \,
 {\rm km/s} \ .
\eeq
We truncate the PS mass function at $\v200 < 40\kms$ (at $z=0$), 
in part to restrict our attention to the range of luminosities where the
disk galaxy population is best characterized observationally and
in part because photoionization heating by the cosmic UV background
is likely to have an important influence on the collapse and
cooling of gas within lower mass halos \citep{tho1996,qui1996,gne2000}.
We also truncate the mass function at $\v200 > 300\kms$, since
disk galaxies with rotation velocities larger than $300\kms$ are rare.
No doubt some disk galaxies do reside within more massive halos,
but these are the common halos of groups that contain multiple
galaxies, and our present calculational approach (which assumes one 
galaxy per halo) cannot be applied to them.

We assume that each halo has a density profile of the form proposed
by Navarro, Frenk, \& White (\citeyear{nav1997}, hereafter NFW; see also 
\citealt{nav1995,nav1996}), which has asymptotic logarithmic slopes
of $-1$ in the central regions (as found by \citealt{dub1991,war1992})
and $-3$ in the outer regions.  The transition between these regimes
occurs at a scale radius $r_s$, and the shape of the profile can 
be characterized by the dimensionless concentration parameter
$c \equiv \rvir/r_s$.  If we adopted the profile form suggested by
\citet{mor1999}, which has an asymptotic inner slope of $-1.5$,
we would generally predict higher circular velocities and flatter
rotation curve slopes for a given galaxy luminosity, though again
we would expect this change to affect mainly the zero points of
relations rather than the overall structure of principal components.

The central densities of halos are connected to the physical density of
the universe at the time that they assemble.  
Concentrations therefore tend to be higher in cosmologies with 
lower $\om0$ or larger mass fluctuation amplitudes, since
these changes shift halo formation to higher redshifts.
In any given cosmology, average halo concentrations are higher for 
less massive halos, which tend to collapse earlier.
\citet{jin1999} and \citet{bul2001} find that there is also significant
scatter in concentrations at a given mass, with a distribution of
concentrations that is roughly log-normal with a scatter
$\sigma_{\ln c}\approx 0.2$.  In all of our calculations, we take
the mean concentration $\bar c$ as a function of halo mass
from Table 2 of \citet{jin1999}.  In some models, we also include
a log-normal scatter in concentrations, taking $\sigma_{\ln c}$
from the same source.  For the star formation calculations 
described in \S\ref{subsec:sfr}, we need to compute the disk
scale length as a function of redshift, which in turn depends
slightly on the halo concentration as a function of redshift.
Based on the numerical results of \citet{bul2001}, we assume that the 
concentration parameter evolves as $c \propto (1+z)^{-1}$.

The angular momentum of the halo is characterized by the dimensionless
spin parameter, $\lambda$, defined as \citep{pee1993}
\beq 
\label{eq:spin}
\lambda= \frac{\Jh \vert\Eh\vert^{1/2}}{G \Mh^{5/2}},
\eeq
\noindent
where $E$ is the halo's total binding energy and $\Jh$ is the halo
angular momentum, which is usually assumed to originate from tidal
torques on the quadrupole moment of the halo near the time of maximum
expansion \citep{pee1969}. Analytic and numerical studies
by several authors \citep{bar1987,war1992,cat1996a,cat1996b} 
find the distribution of spin parameters of collapsed dark matter
halos to be well characterized by a log-normal distribution.
Most authors find that this distribution peaks at $\overline \lambda
\approx 0.05$ with a logarithmic width of $\sigma_{\ln \lambda} \approx 0.5$
(MMW, but see also Dalcanton et al.\ 1997), and that there is little
if any correlation between $\lambda$ and halo mass or initial
peak height \citep{bar1987,ryd1988}.  For most of our calculations, therefore,
we draw the values of $\lambda$ from a log-normal distribution with
these parameters and assume that they are independent of mass and redshift.
We will also consider some models in which all halos have 
$\lambda=\bar\lambda$.
As discussed in \S\ref{subsec:df} below, we will usually follow
the conventional assumption
\citep{fal1980} that the baryons have the same specific angular momentum
as the dark halo and conserve that angular momentum during disk 
formation, so that the value of $\lambda$ effectively determines
the size of the disk.  In some models, we will also allow for
the possibility that baryons lose angular momentum to the dark 
matter during collapse.  

With the assumption that disks have the same specific angular momentum
as their halos, \citet{sye1999} derive a $\lambda$ distribution from 
\citeauthor{mat1996}'s (\citeyear{mat1996})
sample of 2500 late-type spiral galaxies that is in good agreement with
numerical predictions for halos: log-normal with 
$\bar\lambda=0.05$ and $\sigma_{\ln \lambda}=0.36$
(see \citealt{dej2000} for a similar analysis).
Thus, even if the physical process that determines disk sizes
is considerably more complicated than the simple picture of 
collapsed baryons retaining the same specific angular momentum
as the parent halo, the conventional assumption provides a 
good phenomenological description of the observed distribution
of disk sizes.  The fact that the inferred $\lambda$ distribution
is narrower than the predicted one might reflect selection biases
against low surface brightness galaxies (large $\lambda$) and
compact galaxies (small $\lambda$); the latter may become earlier
type galaxies as a result of secular bulge formation.

\begin{figure}
\plotone{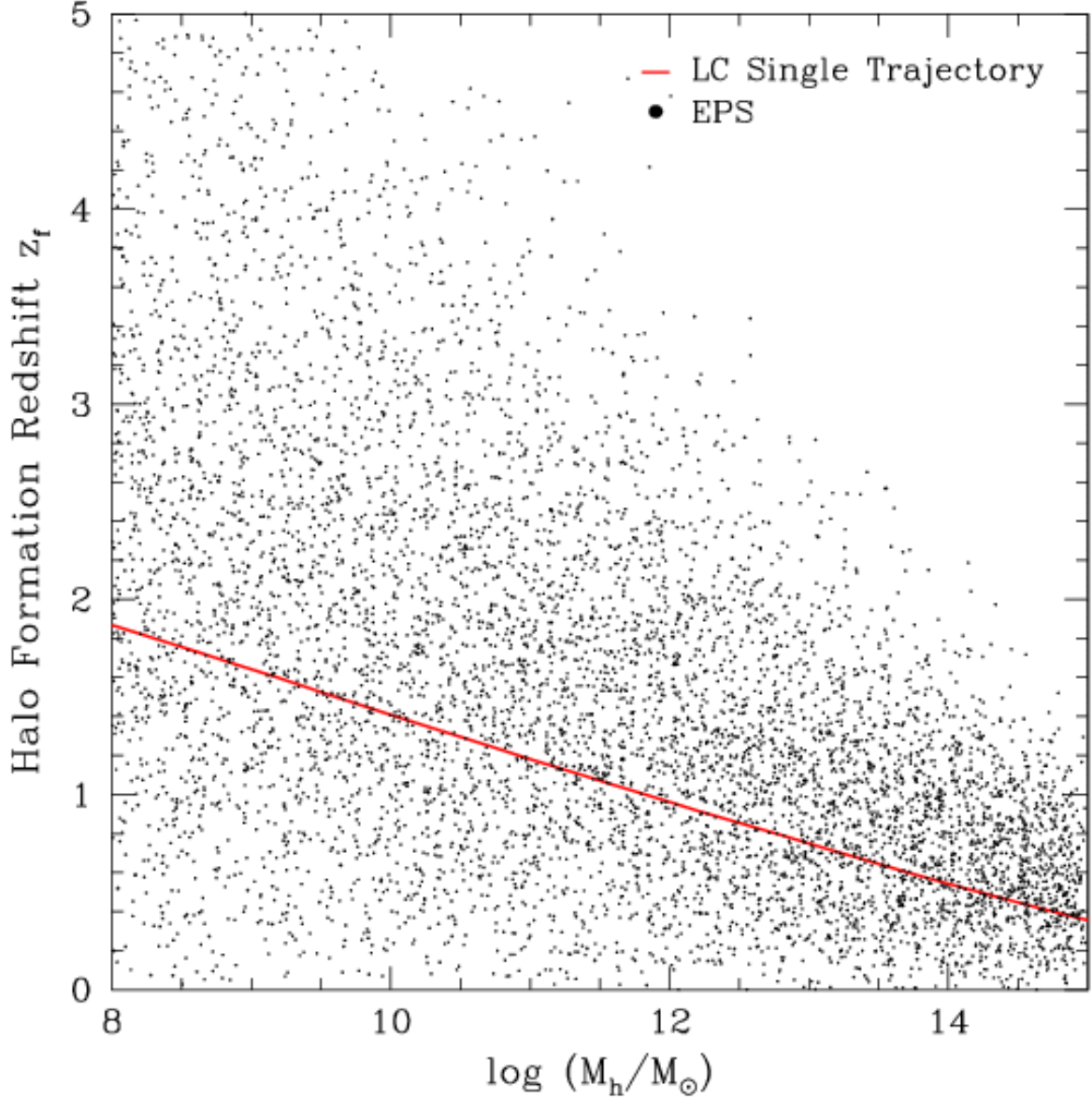}
\caption{Halo formation redshifts, defined as the half-mass 
assembly redshift and calculated by the methods of \cite{lac1993}.
Points show a Monte Carlo realization of the probability distribution
of formation redshifts at different halo masses.  In models where we
suppress the scatter in formation redshifts, we apply the LC
``single trajectory'' relation shown by the solid line.
Our models assume that disk star formation begins at the halo
formation redshift.
\label{fig:zcoll}}
\end{figure}

To calculate the star formation history and spectral evolution
of a galaxy, as described in \S\S\ref{subsec:sfr} and~\ref{subsec:feh}
below, we need to know when the star formation in the galaxy
begins.  We have decided to identify the epoch of initial star
formation with the halo formation redshift, which we take to be
the time when half of the halo's final mass has been assembled.
These formation redshifts can be computed approximately within
the Extended Press-Schechter (EPS) formalism \citep{bon1991,bow1991},
as described by \citet{lac1993} (hereafter LC).
In some of our models, we assume that all halos of a given mass
have the same formation redshift, and in these cases we use the
value implied by equation 2.23 of LC.  In other models, we 
incorporate the full distribution of formation redshifts at fixed 
halo mass predicted by the EPS formalism, using the
prescription in \S 2.5.2 of LC (equation 2.26 in particular;
see the appendix of \citealt{kit1996} for useful approximations).
Figure~\ref{fig:zcoll} shows the relation between formation 
redshift and halo mass obtained from the deterministic relation
(solid line) and the probabilistic description (points).
Clearly there is an overall trend for more massive halos to
collapse at later times.  However, in the probabilistic model,
the scatter in formation redshifts is large in comparison to
this trend.  

Our identification of the start of star formation with the 
half-mass assembly epoch of the halo is, of course, somewhat arbitrary.
Changes to this prescription would shift the ages of the 
stellar populations and hence the predicted colors of our model galaxies.
From the point of view of our PCA calculations, the important
features of this prescription are not its details but the moderate
trend of decreasing formation redshift with increasing halo mass,
and the substantial scatter in formation redshifts in the probabilistic case.
The second of these seems likely to be present in any realistic model,
and the amplitude of scatter predicted by halo formation arguments
seems like a reasonable guess at the actual scatter in
initial star formation epochs.  The reality of the trend with mass
is less obvious, since one could imagine that stars begin to form
in sub-units that assemble into the final galaxy, and that these
sub-units form earlier in systems that are fated to become more massive.
However, our implicit model is that disk stars form from gas that 
settles in the halo only after it has completed most of its growth,
since major mergers would disrupt a pre-existing disk, and within
this picture a trend like that in Figure~\ref{fig:zcoll} is plausible.
In any event, we will find that the formation redshift plays a 
sub-dominant role in determining galaxy spectral properties,
which are more sensitive to the disk surface density, and thus
to the spin parameter $\lambda$.

\subsection{Disk Formation}
\label{subsec:df}

After the gravitational collapse of a dark halo and its associated
baryonic material, gas begins to cool and decouple from the dark matter. 
Since the gas can radiate energy but cannot easily rid itself of
angular momentum, it settles into a centrifugally supported disk,
rotating in the combined gravitational potential of the halo and
the disk itself.  Following MMW, we define $m_d=M_d/M_h$ to be
the ratio of the disk mass to the total mass $M_h$ within the halo 
virial radius.  In a spherical collapse picture, $m_d$ should not
exceed the universal baryon fraction $f_b=\omb0/\om0$; 3-dimensional
hydrodynamic simulations show that cooled gas fractions can sometimes
exceed this limit, but not by a large factor (Gardner et al., in preparation;
Berlind et al., in preparation).  Observational analyses imply
that the mean value of $m_d$ is substantially less than $f_b$,
probably by a factor of two or more, though uncertainties in
disk mass-to-light ratios and cosmological parameters make estimates
quite uncertain (see, e.g., \citealt{fuk1998,sye1999}).
The value of $m_d$ is presumably controlled by
a combination of cooling physics and feedback processes.
If the interaction between cooling and feedback is tightly
self-regulated, we might expect the distribution of $m_d$ to be narrow.
If these processes have large stochastic variations from galaxy
to galaxy, the distribution of $m_d$ could be broad.  We will 
consider both possibilities: models in which $m_d$ is fixed at $0.5 f_b$
and models in which $m_d$ is uniformly distributed in the range $(0,f_b]$.

We assume that the cooled baryons settle into an exponential disk,
with surface density profile
\beq  
\label{eq:expdisk}
\Sigma(r) = \Sigma_0 \exp\left(-r/\Rd\right) = 
\frac{\Md}{2\pi\Rd^2}\exp\left(-r/\Rd\right).
\eeq
\noindent
We compute the scale radius $\Rd$ by requiring centrifugal support.
If the halo were isothermal, the disk had the same specific 
angular momentum as the halo, and the self-gravity of the disk
were negligible, the implied disk scale length would be
$\Rd=\lambda \rvir/\sqrt{2}$ (MMW).  In most of our models,
we assume that the specific angular momentum of the disk is
the same as that of the halo, implying $j_d \equiv J_d/J_h = m_d$.
However, we also consider models with angular momentum loss, 
choosing $j_d$ uniformly in the interval $(0,m_d]$.
Following MMW, we account for the disk's self-gravity
and include the influence of the disk on the dark halo by assuming
that the halo contracts adiabatically as the disk forms.
The adiabatic contraction condition implies that
a halo particle initially at a mean radius $r_i$ will end up at a mean 
radius $r$ such that
\begin{equation}
\label{eq:initialrad}
r_i \, M_i(r_i) = r \, M_f(r),
\end{equation}
where $M_i(r)$ is the initial mass distribution given by the NFW density
profile and 
\begin{equation}
\label{eq:finalmass}
M_f(r) = M_d(r) + (1-m_d)M_i(r_i)
\end{equation}
is the final mass of the system within $r$ \citep{blu1986}. 
$\Md(r)$ is the disk mass within $r$ and is explicitly given by
\begin{equation}
\label{eq:diskmass}
M_d(r) = M_d \biggl[1 - \Bigl(1 + {r \over r_d}\Bigr) \exp(-r/r_d)\biggr].
\end{equation}
Under these conditions, MMW derived the following expressions for the total
disk angular momentum and disk scale length:
\beq
\label{eq:jd}
J_d = \jd \Jh = \frac{2\Md \Rd \v200}{\fr} ,
\eeq
\beq 
\label{eq:rd_nfw}
\Rd = {1\over \sqrt{2}}\left({\jd\over \md}\right)
\lambda \rvir f_c(c)^{-1/2} \fr (\lambda,c,\md,\jd) ,
\eeq
\beq \label{eq:fr_nfw}
\fr (\lambda,c,\md, \jd) 
= 2 \left[\int_0^\infty 
e^{-u} u^2 ~{\vcir(\Rd u)\over\v200}~ du\right]^{-1} \ ,
\eeq
\beq \label{eq:fc_nfw}
f_c(c)={c \over 2}~{1-1/(1+c)^2-2\ln(1+c)/(1+c) \over [c/(1+c)-\ln(1+c)]^2}\ .
\eeq
\noindent
The factors $\fr$ and $f_c$ represent the
necessary corrections to be made to quantities such as the disk scale
length once  the disk self-gravity and halo concentration are taken into
account. In particular, $f_c$ represents the change in binding energy from a
singular isothermal sphere profile to an NFW profile, while the factor
$\fr$ is associated with both the density profile
and the gravitational effects of the disk.
Note that the concentrations in equations~(\ref{eq:rd_nfw})-(\ref{eq:fc_nfw})
depend on redshift, with $c \propto (1+z)^{-1}$ for a given halo.

The rotation curve of the system is obtained by adding
in quadrature the contributions from the halo and the exponential disk.
Before adiabatic contraction, the halo is described by an NFW profile,
and its circular velocity profile can be expressed in terms 
of the concentration parameter:
\beq
\label{eq:vhalo_nfw}
V_c(r) = \v200 \sqrt{\frac{\ln(1+cx)-(cx)/(1+cx)}{x\ln(1+c)-c/(1+c)}}, 
\eeq
\noindent
where $x=r/r_{200}$.
The rotation curve produced by an NFW profile increases
as $r^{1/2}$ at small $r$, reaches a maximum at
$r/\rvir \approx 2/c$, and declines beyond this radius.
The exponential disk contribution to the rotation curve is \citep{fre1970}
\beq
\label{eq:vdisk}
\vd^2(R) = 4\pi G \Sigma_0 \Rd y^2 \left[\calI_0(y) \calK_0(y)
-\calI_1(y) \calK_1(y) \right]\ , ~~~~~~  y\equiv {r\over 2\Rd}\ , 
\eeq
\noindent
where $\calI_n$ and $\calK_n$ are modified Bessel functions of the first
and second kinds. The disk rotational velocity peaks at $r/\Rd \approx
2.2$, where $\vd \approx 0.88 \sqrt{\pi G \Sigma_0 \Rd}$.

To calculate the rotation curve of the final system, we must
obtain $r_d$ and compute the influence of adiabatic contraction
on the halo contribution.  Given values of the halo mass $M_h$,
spin parameter $\lambda$, and concentration $c$, and the disk mass
and angular momentum fractions $m_d$ and $j_d$, we calculate
a first guess at $\Rd$ using equation~(\ref{eq:rd_nfw}) with $\fr=1$.
We then compute the adiabatic contraction caused by a disk with
this scale length, solving for $r_i$ as a function of $r$ using equations
(\ref{eq:initialrad})--(\ref{eq:diskmass}) and thus obtaining $M_f(r)$.
We then calculate the rotational velocity profile $V_c(r)$ and use it
in equation~(\ref{eq:fr_nfw}) to obtain a new value of $f_R$, and hence
a new estimate of $\Rd$ to be used in a second iteration.
This procedure converges rapidly, returning a
value of $\Rd$ to an accuracy better than one percent in a few iterations.

We characterize the amplitude of our model rotation curves by 
$V_{2.2}=V_c(2.2 r_d)$,
the value of the circular velocity at the radius where the disk contribution
peaks.  For observed galaxies, $V_{2.2}$ is usually close to the velocity
used in studies of the Tully-Fisher (\citeyear{tul1977}) relation.
As emphasized by \citet{per1996}, the {\it shape} of the rotation
curve is a valuable diagnostic for the halo profile and the relative
importance of baryon and dark halo contributions.  We characterize
the shape by the logarithmic slope of $V_c(r)$ between 2 and 3
disk scale lengths,
\beq
\label{eq:s23}
\sab \equiv \frac{\Delta\log\vcir}{\Delta\log r} \equiv 
               \frac{\log[V_c(3\Rd)/V_c(2\Rd)]}{\log(3/2)}.
\eeq 
\noindent
It is worth noting that the ``optical radius'' of an exponential 
disk is $R_{\rm opt} = 3.2\Rd$, corresponding to the de Vaucouleurs 
$25\; B$ mag arcsec$^{-2}$ photometric radius for a typical value of
the central surface brightness of $I_0 = 150 L_{\odot}\; {\rm pc}^{-2}$.
Most well observed galaxy rotation curves
extend beyond 3 disk scale lengths \citep{per1996,swa2000}.

\subsection{Star Formation}
\label{subsec:sfr}

The star formation rate (SFR) in observed galaxies is tightly correlated
with the local gas surface density, a relation that can be approximately
characterized by a power law \citep{sch1959} with a minimum threshold
for active star formation \citep{ken1989,ken1998}.
\cite{ken1998} finds that the star formation rates of disk galaxies
are well described by the relation
\beq
\label{eq:ken}
\Sigma_{\rm SFR} = A
\left({\Sigma_{\rm gas}\over M_\odot {\rm pc}^{-2}}\right)^n ,
\eeq
\noindent
where $\Sigma_{\rm SFR}$ and $\Sigma_{gas}$ represent the disk-averaged
SFR and gas densities, $A=(2.5\pm 0.7) \times 10^{-4} 
{\rm \msun yr^{-1} kpc^{-2}}$, and $n=1.4\pm 0.15$.

To derive star formation histories of our disks, we draw on the
analytic results of \cite{hea1999}, and we thus adopt the model
of disk evolution that is implicit in their calculations:
an exponential gas disk forms at initial redshift $z_i$, 
which we identify with the halo formation redshift $z_f$, and
this gas is converted into stars at the rate implied by 
equation~(\ref{eq:ken}). The scale length of the disk is determined by
the values of $\lambda$ and the virial radius.
With this model, \citet{hea1999}
derived the time evolution of the gas surface 
density in terms of the initial gas surface density $\Sigma_{i,{\rm gas}}$:
\beq
\label{eq:sigmat}
\Sigma_{\rm gas}(t) = \left(\Sigma_{i,{\rm gas}}^{-0.4}+0.4 Bt\right)^{-2.5} \ ,
\eeq
\noindent
where $B=9.5 \times 10^{-17}$ in SI units and $t$ is the time
elapsed since the initial redshift $z_i$.
Integrating equation~(\ref{eq:sigmat}) over the entire exponential
disk yields the overall star formation rate and remaining gas mass
as a function of time:
\bea
\label{eq:sfr}
\dot{M}_{\star}(t) & = & \frac{50 \pi B \Rd^2 \Sigma_{0}^{1.4}}{49}
\phantom{I}_3 F_2\left(3.5,3.5,3.5;4.5,4.5;-a\right)\ M_\odot {\rm yr}^{-1},\\
M_g(t) & = & {2\pi \Rd^2 \Sigma_{0}} \phantom{I}_3
F_2\left(2.5,2.5,2.5;3.5,3.5;-a\right)\ M_\odot,
\eea
\noindent
where $\!\phantom{I}_3 F_2$ is a generalized hypergeometric function
\citep{gra1980}, $\Rd$ is the disk scale length, and 
$\Sigma_{0}=M_d/2\pi r_d^2$ is the central gas surface density.  
The quantity $a$ is a time-like variable defined by 
\beq 
\label{eq:avar}
a = 1.06 h_z^{0.4}\left({t\over \gyr}\right) \left({\v200\over
250 \kms}\right)^{0.4}\left({\lambda\over
0.05}\right)^{-0.8}\left({m_d\over 0.05}\right)^{0.4} 
\left({j_d\over m_d}\right)^{-0.8} f_c^{\ 0.4}\  
\fr^{-0.8},
\eeq
\noindent
where $h_z \equiv H(z)/100$ km s$^{-1}$ Mpc$^{-1}$.
\citet{hea1999} assumed $j_d=m_d$, an isothermal halo, and no 
disk self-gravity or adiabatic contraction of the halo.  
Equation~(\ref{eq:avar}) differs from their equation (9) by
the inclusion of the last three factors, which account, respectively,
for angular momentum loss, NFW halo profiles, and the influence
of self-gravity and adiabatic contraction on the disk scale length.
\citet{hea1999} use the values of $r_d$, $\Sigma_0$, and $h_z$
determined at the formation redshift, thus implicitly assuming
that the disk forms with its full mass $M_d=m_d M_h$ at $z=z_f$
with size proportional to the halo virial radius at that redshift.
In order to approximately treat the growth of the disk over time,
we use the time-dependent values
of $\Rd$, $\Sigma_0$, $h_z$, and $c$ in equations
(\ref{eq:sfr})-(\ref{eq:avar}), assuming that the halo evolves
at constant circular velocity $V_{200}$ and thus has
$M_d(z)=m_d M_h(z) \propto H^{-1}(z)$ and $r_d \propto H^{-1}(z)$
(see eq.~\ref{eq:massh}).
This treatment is intermediate between assuming that the disk size is
determined at $z_f$ and assuming that it is determined at $z=0$.  

This calculation of a disk's star formation history is clearly
idealized and approximate.  However, it is two basic features
of this prescription that are crucial to our results.
The first is that the {\it gas surface density} is the primary
physical parameter that controls the speed at which gas converts
into stars.  The second is that the formation redshift of the 
halo determines the initiation of star formation.  

As an alternative to our standard models, we also create
realizations of the galaxy population in which the gas surface density
does {\it not} drive the SFR.
For these models, we assume an exponentially decaying star formation rate 
with the decay time $\tau$ drawn from a uniform distribution in the interval
$[1,9] {\rm\,Gyr}$, independent of the surface density.
The disk mass again grows as $M_d(z)=m_d M_h(z)$, starting at the 
halo formation redshift, and gas added to the disk at time $t_a$
is converted into stars at a rate ${\rm SFR} \propto \exp[-(t-t_a)/\tau]$,
with the constant of proportionality chosen so that all of the gas
would be consumed as $t \longrightarrow \infty$.
In these exponential decline models, the
star formation history is decoupled from structural quantities like the
spin parameter $\lam$ and the disk mass fraction $\md$.
They provide valuable insight into the origin of the principal
component structure of our standard models and show how that
structure would change if star formation is not tightly coupled
to the surface density over the full history of the disk.

\subsection{Spectral and Chemical Evolution}
\label{subsec:feh}

Given the star formation histories calculated above, we
compute the spectral evolution of the stellar populations
using the current version of the \citet{bru2001} spectrophotometric
population synthesis (SPS) code. 
These calculations return observational quantities that 
play a central role in our principal component analysis,
such as broad band magnitudes and colors, 
stellar mass-to-light ratios, and the strength of
the 4000\AA \ spectral break. 
We implement a fully consistent chemical enrichment model
that makes use of the latest \citet{bru2001} SPS models and
of metallicity dependent lifetimes and yields.  
We describe the details of this chemical enrichment model
in the Appendix.

In all of our evolutionary calculations, we assume that the
stellar initial mass function (IMF) has the form proposed
by \citet{sal1955}, with a logarithmic slope $x = 1.35$ over
the mass range $m_l = 0.1 M_\odot$ to $m_u = 125 M_\odot$. 
Changes to the lower cutoff (or to the form at low masses) 
would alter the stellar mass-to-light ratios but have little
impact on spectral properties, since low mass main sequence stars 
contribute much of the mass but little of the light for a typical 
stellar population.  Changes to the upper cutoff would influence
the enrichment history, since the most massive stars contribute
significantly to heavy element production.  Changes to the slope
in the regime $\sim 1-10 M_\odot$ would have the most important
impact on spectral properties, though even these changes would
tend to shift colors in a coherent way without altering the
degree of correlation between colors and other galaxy properties.
From the point of view of PCA predictions, the key assumption
is not the particular form of the IMF but that it is universal,
at least when averaged over a galaxy's history.

\subsection{Summary: Inputs and Outputs}
\label{subsec:parameters}

In our calculations, an individual galaxy is defined by six input parameters, 
or seven in the case where the star formation timescale is chosen independently
of the surface density.  The ten galaxy formation models that we examine
in \S\ref{sec:results} differ in the distributions of these input parameters:
which ones are held fixed at typical values and which ones are allowed to vary 
and therefore play a role in governing the distribution of galaxy properties.

The input parameters (listed in Table~\ref{tab:input}) 
are the halo mass $M_h$,
the halo concentration parameter $c$,
the halo spin parameter $\lambda$,
the halo formation redshift $z_f$, the ratio of disk mass to halo mass
$m_d=M_d/M_h$, the ratio of disk angular momentum to halo angular
momentum $j_d=J_d/J_h$, and (in some models) the $e$-folding time
of the star formation rate $\tau$.
The values of $M_h$ and $m_d$ determine the disk mass.
The disk scale length is determined by the condition of centrifugal
support in the combined potential well of the disk and the adiabatically
contracted halo, so it depends on $M_h$ (which fixes the halo virial radius),
$\lambda$, and $j_d/m_d$, and, more weakly, on $c$ and $m_d$.
Given the size and mass of the disk and the profile of the adiabatically
contracted halo, we compute the rotation curve as described in 
\S\ref{subsec:df}.  The size and mass of the disk also determine the
surface density profile, which we use to compute the star formation
history as described in \S\ref{subsec:sfr}, except in models where
we choose the timescale $\tau$ independently of surface density.
The halo formation redshift $z_f$ also influences the star formation
history by defining the initial epoch of star formation.
Given the star formation history, we compute spectral and chemical
evolution of the galaxy as described in \S\ref{subsec:feh} and
the Appendix.

The results of PCA will depend to a large
extent on what observables we choose to incorporate into the analysis.
Of the large number of observables that can be computed by our code,
we have chosen to work with the set listed in Table~\ref{tab:output}.
These fall roughly into two categories, quantities that describe
the scale or structure of the galaxy and quantities that describe
the shape of the spectral energy distribution (SED).  The first category
includes the I-band disk luminosity $\Ldi$, the exponential scale
length $\Rd$, the circular velocity at 2.2 disk scale lengths $\vdmax$,
the rotation curve slope $\s23$, and the I-band central surface
brightness $\mu_{0,I}$.  Since we assume that the disk is exponential,
the central surface brightness is determined by the luminosity and the
scale length, but it proves helpful to treat it as a separate observable
because of its direct connection (in Schmidt-law models) to the 
star formation history.
We also include the disk stellar mass $M_\star$ and the I-band stellar
mass-to-light ratio $(M/L)_I$ as observables.  In most galaxies it is not
possible to measure the stellar mass dynamically, but one can infer
a galaxy's stellar mass-to-light ratio by population synthesis modeling 
of the SED and multiply by the luminosity
to infer a stellar mass, albeit with some uncertainties.
While $M_\star$ is determined by $\Ldi$ and $(M/L)_I$ in such an analysis, 
it is still helpful
to treat it as a separate observable in PCA to distinguish between the effects
of disk mass and the effects of stellar population on the I-band luminosity.

For the SED observables we have selected three broad-band colors
that probe somewhat different features of the stellar population, $(U-B)$,
$(B-V)$, and $(V-K)$, and the strength of the 4000\AA\ break $B_{4000}$.
We also include the ``birth parameter'' $b$ \citep{sca1986}, the 
ratio of the galaxy's current star formation rate to its
time-averaged star formation rate:
\beq
b(t) = \frac{{\rm SFR}(t)}{\langle {\rm SFR} \rangle} \approx
\frac{{\rm SFR}(t) \times t}{M_{\star}(t)} \ .
\eeq
As shown by \citet{bru1993}, the SED of a model galaxy is usually
highly correlated with $b$, even for different overall star 
formation histories.  It therefore serves us both as a scaled 
characterization of ongoing star formation and as an SED quantity.
Our final observable is the mean metallicity $[{\rm Fe/H}]$.

\section{Principal Component Analysis}
\label{sec:pca}
Principal Component Analysis is among the oldest and
best known of the techniques of multivariate analysis. It was first
introduced by \citet{pea1901} and developed independently by
\citet{hot1933}. The central idea behind PCA is to reduce
the dimensionality of a data set in which there are a large number of
interrelated variables, while retaining as much as possible of the
variation of the whole data set. This reduction is achieved by
transforming to a new set of variables, the principal components
(hereafter PCs), which
are uncorrelated and are ordered so that the first {\it few} retain most
of the variation present in {\it all} the original variables. 

Algebraically, PCs are particular linear combinations
of the $p$ variables. Geometrically, these linear
combinations represent the selection of a new coordinate system obtained
by rotating the original system with ${\bf x} = x_1$, $x_2 \ldots x_p$ as the
coordinate axes. The new axes represent the direction of maximum
variability and provide a simpler description of the covariance structure.

We seek to find some new variables {\bf $\xi$} = $\xi_1$, $\xi_2 \ldots
\xi_p$ which are linear functions of the $x$'s but are themselves
uncorrelated. In fact, we look for $p^2$ constants ${\bf l} = l_{ij}
(i,j=1,\ldots,p)$ such that 
\beq
{\bf \xi = l \  x} \ ,
\eeq
with the orthogonality condition ${\bf ll^{\prime} = I}$,
where {\bf I} is the identity matrix. It can be shown \citep{mur1987}
that if {$\bf \Sigma$} is the variance-covariance matrix of the
original ${\bf x}$ vector of $p$ variables, then the axis along which
the variance is maximal is the eigenvector ${\bf e_1}$ of the matrix
equation
\beq
{\bf \Sigma e_1} = \lambda_1 {\bf e_1},
\eeq
where $\lambda_1$ is the largest eigenvalue, which is in fact the
variance along the new axis. The other eigenvalues obey similar
equations. The total population variance is then
\beq
\sigma_{11} + \sigma_{22} + \cdots \sigma_{pp} = \lambda_1 + \lambda_2 +
\cdots \lambda_p ,
\eeq
and hence the proportion of total variance explained by the $k$th
principal component is
\beq
\label{eq:variance}
\frac{\lambda_k}{\lambda_1 + \lambda_2 +
\cdots \lambda_p} \ \ \ \ \ \ \ \ \ \ \ k=1,2,\ldots,p \ . \\
\eeq
Thus, the first axis accounts for as much of the total
variance as possible; the second axis accounts for as much of the
remaining variance as possible while being uncorrelated with the first
axis; the third axis accounts for as much of the total variance
remaining after that accounted for by the first two axes, while being
uncorrelated with either, and so on.
If most of the total population variance can be attributed to the first
few components, then these components can replace the original $p$
variables without much loss of information. The magnitude of each of the
$e_{ki}$ eigenvector coefficients measures the importance of the $k$th
variable of the $i$th principal component, irrespective of other variables.

It is common (and advisable) to remove any variance introduced only
by the widely differing dynamic ranges in variable measures. This is
particularly important in our own case when we perform PCA on quantities
with measurement units that are not commensurate (see Table
\ref{tab:output}). We will therefore 
transform our original variables into a new set with zero mean and unit
variance. In matrix notation:
\beq
{\bf z} = ({\bf V^{1/2}})^{-1} ({\bf x - \mu})
\eeq
where ${\bf V^{1/2}}$ is the diagonal standard deviation matrix.
The PCs of ${\bf z}$ can then be obtained from the
eigenvectors of the $correlation$ matrix ${\bf \Theta} = ({\bf
V^{1/2}) \Sigma (V^{1/2})^{-1}}$ of ${\bf x}$. All
our previous results still apply, with some simplification since the
variance of each $z_i$ is now unity. Furthermore, since the total
variance is now $p$, 
equation~(\ref{eq:variance}) becomes:
\beq
\frac{\lambda_k}{p} \ \ \ \ \ \ \ \ \ \ \ k=1,2,\ldots,p \ .
\eeq
The PCs derived from ${\bf
\Sigma}$ are, in general, not the same as the ones derived from ${\bf
\Theta}$.  

In addition to scaling input variables to unit variance, for quantities
that characterize size, mass, or velocity scales we take logarithms of
observables prior to applying PCA, as indicated by the notations in
Table~\ref{tab:output}.  Specifically, we use $\log \Ldi$, $\log\Rd$,
$\log\vdmax$ and $\log M_\star$ (the central surface brightness
$\mu_{0,I}$ is in magnitude units and therefore already
logarithmic). Logarithmic measures reduce the very large dynamic range
in these quantities, and the observed power-law correlations between
these  quantities translate into linear correlations in their
logarithms, making the logarithmic measures more appropriate to the
linear  analysis of PCA.

Once the PCs have been determined, attention must focus
on the relationship of the PCs to the original
variables. In order to do so, we first need to address 
the problem of the number of PCs to retain for further analysis.
One rule for determining how many PCs to retain
was proposed by \citet{kai1960}. The idea behind this rule is
that if all elements of ${\bf x}$ are independent, then the PCs are the
same as the original variables, and all have unit variances. Thus any PC
with variance less than one contains less information than one of the
original variables, and so it is not worth retaining. It can be
argued that Kaiser's rule retains too few variables in the case where
there are a few variables that are more-or-less independent of all
others. These variables, in fact, will have small coefficients in
some of the PCs, but will dominate others, whose variance
will be close to one. Since these variables provide independent
information from the other variables it would be unwise to ignore
them. Based on simulations, \citet{jol1972} proposed to lower the
variance threshold to 0.7. 

The galaxy formation models below offer a natural, physically
motivated choice for the number of PCs to retain in our 
theoretical analyses: the number of physically significant
PCs should equal the number of independently varying input
parameters in the model.  In every case but one, we find that this choice
also corresponds to the number of PCs above a variance threshold of 0.7.

\section{The Principal Component Structure of Model Galaxy Populations}
\label{sec:results}

Table~\ref{tab:models} lists the input parameter values (or distributions)
for the ten models of galaxy formation that we study in this section.
For each model, we construct a realization of $\sim 500$ disk galaxies
evolved to $z_0=0$.  We compute the quantities listed in 
Table~\ref{tab:output} for each galaxy and apply PCA to these
500 vectors of 13 observables.
Models 1 and 2.1-2.5 are deliberately simplified models, with the
expected physical variation in one or more input parameters
suppressed so that we can isolate the physical effects of others.
We discuss these models in \S\ref{subsec:subminimal}.  
Model 3 represents our baseline model of galaxy formation, which 
incorporates variations in $\Mh$, $\lambda$, and $z_f$.
We discuss this model in \S\ref{subsec:minimalplus}.
Models 4.1-4.3 are extended models, where some additional physical
variation is added to the baseline model.  We designate those input
parameters that vary (i.e., are drawn from a probability distribution
rather than fixed to a typical value) as a model's {\it control parameters}.
Thus, our models of galaxy formation differ in which control parameters
govern the variations in galaxies' observable properties.

\subsection{Simplified Models}
\label{subsec:subminimal}

\begin{figure}
\plotone{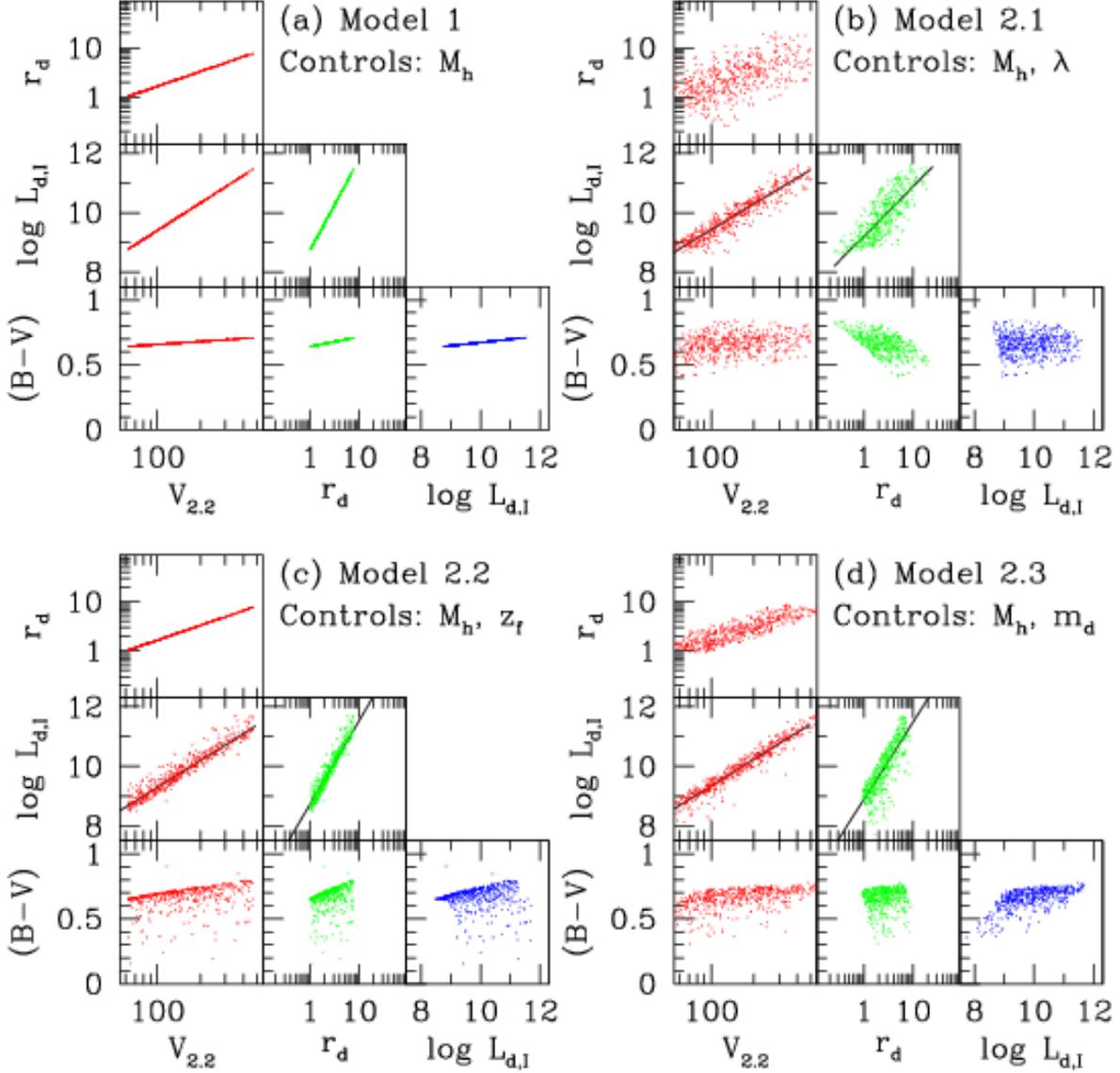}
\caption{
Bivariate correlations among 
disk rotation velocity (at $2.2\Rd$), scale length, 
I-band luminosity, and $\bv$ color in models with one or two
control parameters.  Galaxies in Model 1 form a 1-parameter
family, controlled entirely by halo mass $\Mh$, so there is no
scatter in the correlations.  Models 2.1, 2.2, and 2.3 incorporate
scatter in the spin parameter $\lambda$, formation redshift $z_f$,
and disk mass fraction $m_d$, respectively, which adds scatter
to the bivariate relations and in some cases changes their slopes.
Lines in the $L-V$ and $L-\Rd$ panels show least squares fits to 
the data points.
\label{fig:matrix4subfit}}
\end{figure}

Figure~\ref{fig:matrix4subfit} presents bivariate correlations
among four observables, $\vdmax$, $\Rd$, $\Ldi$, and $\bv$,
for four of our simplified models.  Model 1, shown 
in Figure~\ref{fig:matrix4subfit}a, has halo mass $\Mh$ as the
only control parameter.  Since the galaxies in this model form
a 1-parameter family, the correlations among the 
observables are scatter free, and they are nearly linear 
in these logarithmic plots.
More massive halos host galaxies that have larger $\Ldi$ because
of their larger disk masses, larger $\Rd$ because of their
larger virial radii, and larger $\vdmax$ because of their
larger disk and halo masses.  
The Luminosity-Velocity (L-V, a.k.a. Tully-Fisher) relation
is steeper than the Radius-Velocity (R-V) relation, since
$M_d \propto M_h$ while $\Rd \sim \lambda\rvir \propto M_h^{1/3}$. 
More massive halos host redder galaxies because they have higher surface
densities, roughly $M_d/\Rd^2 \propto \Mh^{1/3}$, and they thus promote
more rapid star formation; this effect wins out over the 
competing trend (see Figure~\ref{fig:zcoll}) of lower $z_f$ at 
higher $\Mh$.

When we add the expected log-normal distribution of spin parameters,
we obtain the model depicted in Figure~\ref{fig:matrix4subfit}b.
While there is still a tendency for more massive halos to host
galaxies with larger scale lengths, the broad distribution of disk
angular momenta at fixed mass produces substantial scatter in the R-V
and L-R relations, though the mean trends are similar to those
of Model 1.  The dispersion in $\lambda$ also adds substantial
scatter to the L-V relation because, for a given halo mass,
more compact disks have larger self-gravity and produce greater
adiabatic contraction of the halo, boosting $\vdmax$.  
The dispersion in $\lambda$ also produces a large dispersion
in colors, because larger disks have lower surface densities
and therefore, with our Schmidt law prescription for star formation,
convert their gas into stars more slowly.  
In Model 1, larger disks always come from more massive halos and
therefore have somewhat higher surface densities, making them redder.
In Model 2.1, with a wide range of disk sizes at fixed mass, the
larger disks are typically of lower surface density, and therefore bluer.

Figure~\ref{fig:matrix4subfit}c shows Model 2.2, with  
the spin parameter fixed to $\lambda=0.05$ and formation redshift
varying as predicted by the EPS formalism (points in Figure~\ref{fig:zcoll}).
The R-V relation is identical to that of
Model 1, where only halo mass varies, because the formation redshift does not
affect the disk scale length. 
The L-V and L-R relations, on the other hand,
pick up a small degree of scatter from $M/L$ variations. 
The color correlations display a ridge line of red galaxies that is close to 
the locus of the Model 1 correlations, but low formation redshifts
produce scatter towards blue colors.

Figure~\ref{fig:matrix4subfit}d shows Model 2.3, in which the disk mass
fraction $m_d$ plays the role of the second control parameter.
(Note that we keep $j_d/m_d$ fixed, so that the disk's specific 
angular momentum is still determined by $\lambda$.)
Variations in $m_d$ produce scatter in the R-V relation mainly because
of disk and adiabatic contraction effects on $\vdmax$; self-gravity
and adiabatic contraction also have a modest impact on $\Rd$ itself.
Most remarkably, the core of the L-V relation, and to a lesser extent
the L-R relation, remains quite tight, similar to the Model 1 locus.
Naively, one might expect changes in $m_d$ to simply add scatter in
$\Ldi$ at a given $\vdmax$.  However, as emphasized by 
\cite{mo2000} and \cite{nav2000},
an increase in $\md$ produces increased self-gravity and adiabatic
contraction that boost $\vdmax$, with the result that points shift
{\it along} the L-V relation rather than across it.  Galaxies with 
low $\md$ scatter to blue colors because of their lower surface densities
and correspondingly slow star formation.  The ridge line of color is 
redder than the locus of Model 1 because the maximum value of $\md$
is the baryon fraction $\md=f_b=0.085$ rather than $\md=0.5f_b$ as
used in Model 1.

The log-normal distribution of $\lambda$ is very broad, with the
result that Model 2.1 exhibits much larger scatter in bivariate
relations than Models 2.2 or 2.3, though these have some large 
outliers in cases where $z_f$ or $\md$ is close to zero.  The spread
in disk sizes at fixed mass reverses the correlation between 
color and scale length relative to the other models, since 
$\lambda$ ``wins'' over $\Mh$ as the
determining factor that sets the surface density and thus the star 
formation timescale.
While $m_d$ and $z_f$ also affect surface density at fixed halo mass, the
distributions of these parameters are not broad enough to overwhelm the
trend of increasing surface density with increasing halo mass. 
The dominance of $\lambda$ as a control parameter and the influence
on galaxy SED through the surface density dependence of the 
Schmidt law anticipate some of the key features we will find for
the baseline model in \S\ref{subsec:minimalplus}.

\begin{figure}
\plotone{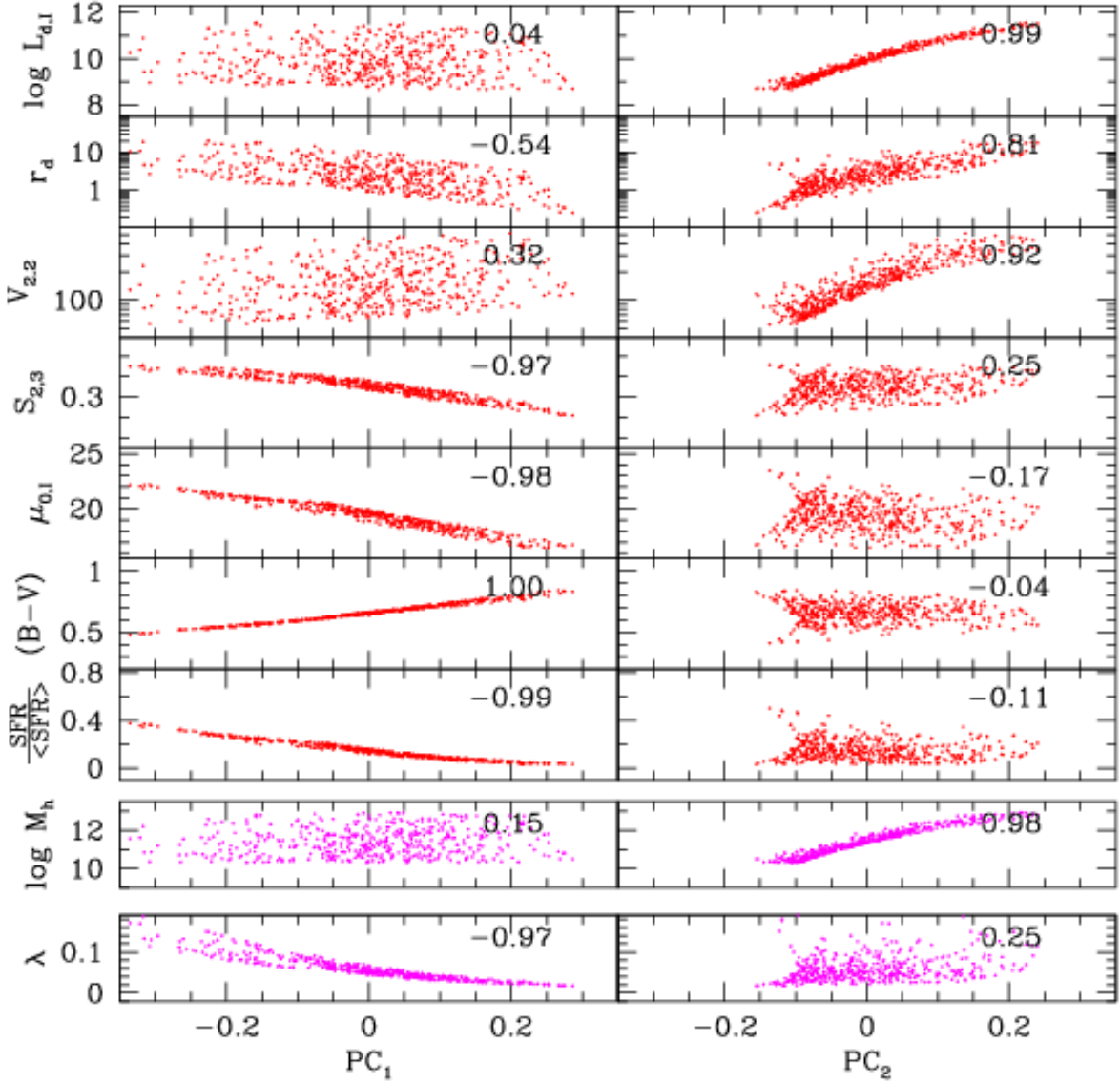}
\caption{
Correlations of observables and control parameters with the first (left
column) and second (right column) principal components of Model 2.1,
which has control parameters $\Mh$ and $\lambda$.  In each panel,
points represent the 500 galaxies in the model realization, and
numbers indicate the Spearman rank correlation coefficient with the 
corresponding PC.  The top rows show correlations
for seven of the 13 observables that enter the PCA.  The bottom
two rows show correlations of the control parameters, which do not 
enter the PCA itself but which drive the principal component structure.
\label{fig:lcdm.scr.model2}}   
\end{figure}

Figure~\ref{fig:matrix4subfit} shows only bivariate relations, and
it incorporates only four of the 13 observables that we compute
for each galaxy.  PCA is an ideal tool for revealing the information
in multi-variate correlations of large numbers of observable quantities.
We apply PCA to the 13 observables of the $\sim 500$ galaxies
of each of our models.  Figure~\ref{fig:lcdm.scr.model2} 
shows the projection of seven of these observables onto the 
PCs recovered for Model 2.1. 
In addition to the luminosity, scale length, rotation speed, and $\bv$
color, we include three other quantities that have high correlations
with the derived PCs: the rotation curve slope $\s23$, the 
I-band central surface brightness $\mu_{0,I}$, and the 
birth parameter $b = \birth$. 

Each panel shows points for each of the model galaxies,
displaying correlations with the first principal component
($\pc_1$) in the left-hand column and the second principal
component ($\pc_2$) in the right-hand column.
The bottom two rows of Figure~\ref{fig:lcdm.scr.model2} show the
correlations of the model control parameters, $\Mh$ and $\lambda$,
with the derived PCs.  These quantities do not enter the PCA
itself because they are not observable, but they are correlated
with the PCs because they govern the correlations of the observables.
Each panel of Figure~\ref{fig:lcdm.scr.model2} lists the Spearman rank
correlation coefficient of the plotted quantity with the corresponding PC. 
Because these correlation coefficients depend only on rank with
respect to other model galaxies, they are not affected by curvature
provided that the correlation remains monotonic, and it does not matter
whether we use linear or logarithmic, normalized or unnormalized
variables (though these choices do matter for the computation of
the PCs in the first place).

Since Model 2.1 has only two control parameters, there are, not
surprisingly, only two significant principal components.
>From Figure~\ref{fig:lcdm.scr.model2} it is evident that $\pc_1$
is predominantly a measure of SED shape, represented here
by $\bv$ and the birth parameter, but by more observables in the
PCA (see Table~\ref{tab:output}).
$\pc_1$ is strongly anti-correlated with the central surface brightness
$\mu_{0,I}$ because of the strong link between star formation history
and surface density provided by the Schmidt law. Note that,
because we use units of mag arcsec$^{-2}$ for $\mu_{0,I}$,
high values correspond to low surface densities, and hence to 
slow gas consumption and blue color.
 
$\pc_1$ is anti-correlated with the scale length $\Rd$, but much
less directly, since, for example, massive disks (in high $\Mh$ halos)
can have large $\Rd$ and still have high surface densities.
Because the baryon effects on the rotation curve are stronger in 
high surface density disks, $\pc_1$ is also strongly (anti-)correlated
with $\s23$ and somewhat correlated with $\vdmax$.

$\pc_2$ is predominantly a measure of overall scale, being
tightly correlated with $\Ldi$ and $\vdmax$ and slightly less so with $\Rd$.
$\pc_2$ is orthogonal to $\pc_1$ by construction, so it is almost 
entirely uncorrelated with $\s23$, $\mu_{0,I}$, $\bv$, and the birth parameter,
which have near perfect correlations with $\pc_1$.
In Model 2.1, $\pc_1$ recovers 65\% of sample variance and $\pc_2$ 27\%.
However, the relative strength
of the PCs is determined largely by our choice of observables.
We incorporate several different measures of
SED shape into the PCA, and these are tightly coupled to each other by
their similar dependence on star formation history. Thus $\pc_1$ accounts for
much of the total variance. If we use only a single color to represent
SED shape, then scale quantities make a much more important contribution
to $\pc_1$, though they also remain correlated with $\pc_2$.

The bottom rows of Figure~\ref{fig:lcdm.scr.model2} demonstrate a
very clean physical division between the PCs of this model.
$\pc_1$ is driven almost entirely by variations in $\lambda$, which
govern the disk surface density, and therefore the star formation
history, and therefore the SED.
$\pc_2$ is driven almost entirely by halo mass, which determines
the disk luminosity and plays the dominant role in governing $\vdmax$
and $\Rd$.

\begin{figure}
\plotone{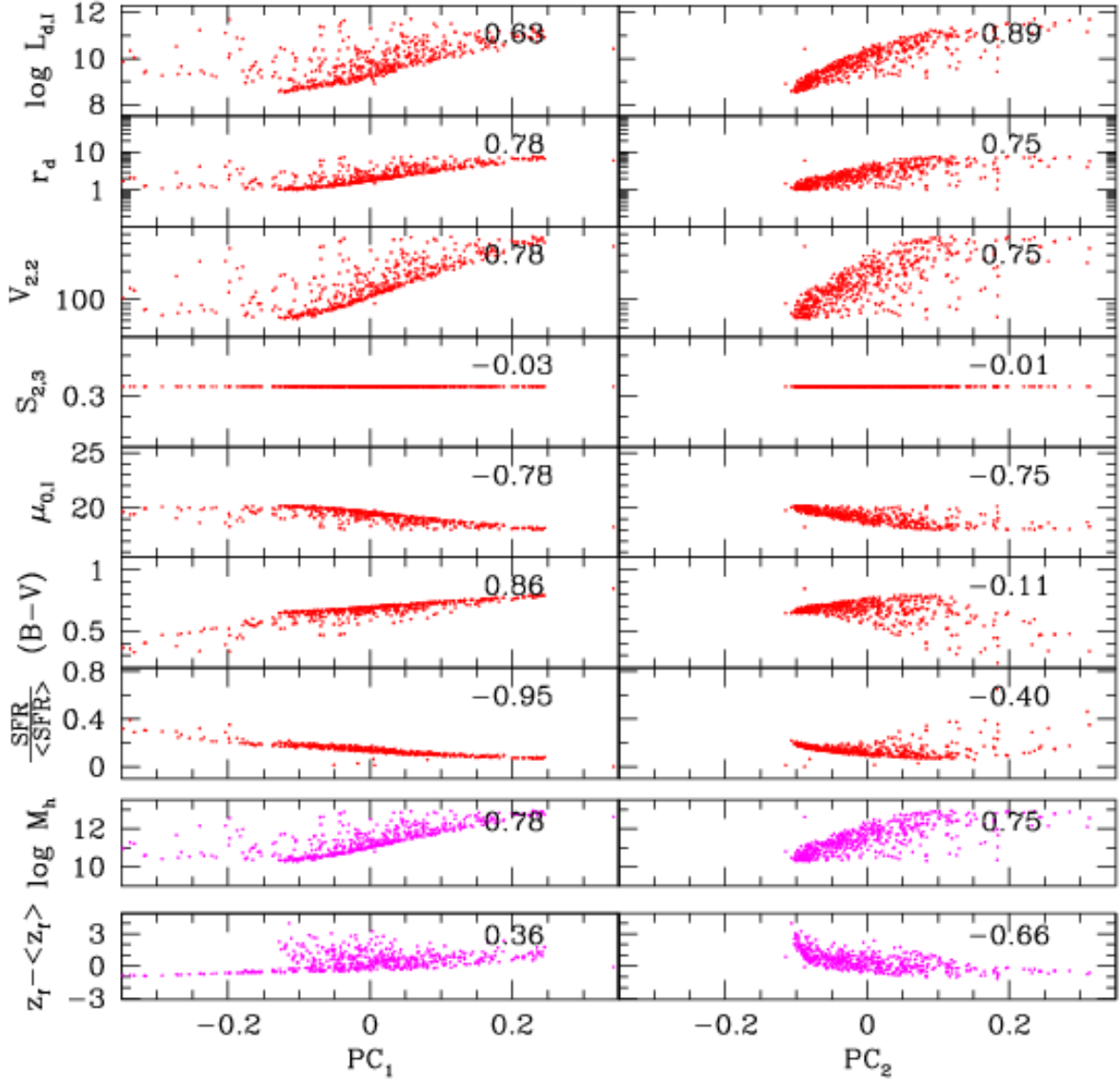}
\caption{
Same as Fig.~\ref{fig:lcdm.scr.model2}, but for Model 2.2, with
control parameters $\Mh$ and $z_f$.
Since $z_f$ varies systematically with halo mass, we subtract
off the mean value $\langle z_f \rangle$ for the galaxy's $\Mh$
in order to isolate variations about the mean trend.
\label{fig:lcdm.scr.model3}}   
\end{figure}

Figure~\ref{fig:lcdm.scr.model3} shows the PC correlations for Model 2.2, 
in which $z_f$ replaces $\lambda$ as the second control parameter.
$\pc_1$ is still strongly correlated with SED quantities ($\bv$ and
birth parameter in this plot), which are again tightly coupled
to central surface brightness.  However, with $\lambda$ variation
suppressed, halo mass becomes the physical driver of surface density
variations and hence the primary determinant of SED shape.
Since halo mass also drives scale quantities like $\vdmax$, $\Ldi$,
and $\Rd$, these also become strong components of $\pc_1$, much more
so than in Figure~\ref{fig:lcdm.scr.model2}.  Formation redshift
plays a minor role in driving $\pc_1$, mainly as a result of galaxies
with very low formation redshifts that have very blue colors.
$\pc_2$ is again composed largely of scale quantities, but the
correlations are weaker than those of Model 2.1 because the correlation
of these quantities with SED properties has been absorbed into $\pc_1$.
$\Mh$ and $z_f$ are both correlated with $\pc_2$, with
comparable strength.  With our full set of observables,
$\pc_1$ recovers more than 50\% of the
sample variance for this model, while $\pc_2$ recovers only 25\%.

\begin{figure}
\plotone{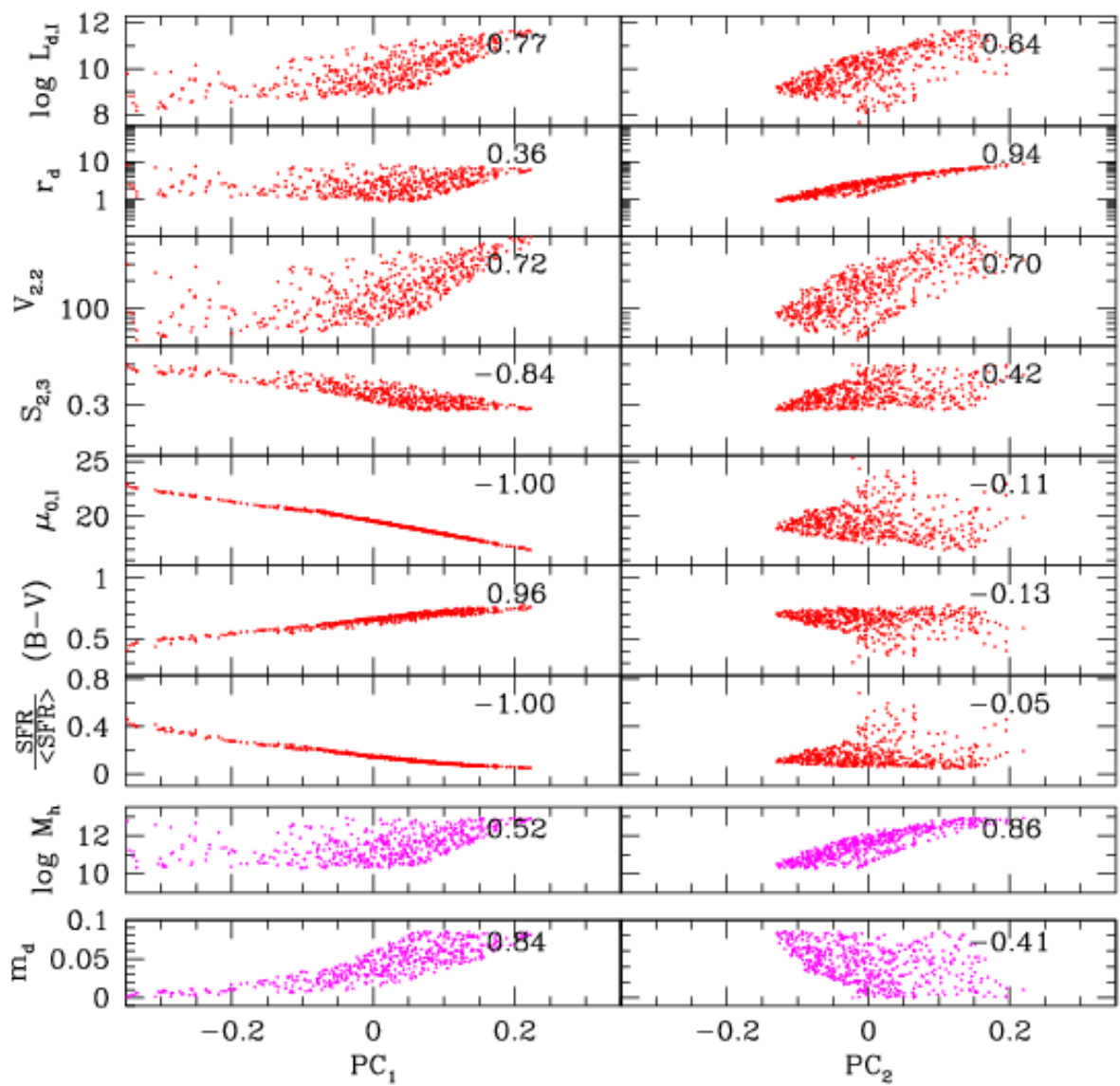}
\caption{
Same as Fig.~\ref{fig:lcdm.scr.model2}, but for Model 2.3, with
control parameters $\Mh$ and $m_d$.
\label{fig:lcdm.scr.model5}}   
\end{figure}

For Model 2.3 (Figure~\ref{fig:lcdm.scr.model5}), 
$\pc_1$ is once again dominated by SED measures
and surface brightness. The main physical driver in this case is the
disk mass fraction $m_d$, since there are no $\lambda$ variations to
induce a change in surface density, but halo mass also has a
significant influence. 
The correlations of control parameters with $\pc_1$ are defined
largely by ``exclusion zones'' --- galaxies cannot attain high surface 
densities and red colors if they have low $\md$ or low $\Mh$,
and they cannot have low surface density and blue colors if they
have high $\md$.  Because $\md$ has a direct impact on disk mass
and determines the baryon contribution to the rotation curve,
$\Ldi$, $\vdmax$, and $\s23$ are substantially correlated
with $\pc_1$; the absence of red galaxies with low $\Mh$ adds to
these correlations and produces a mild correlation with $\Rd$.
$\pc_2$ is again correlated with scale measures, most strongly
with $\Rd$, which, in contrast to $\Ldi$ and $\vdmax$, is only
minimally affected by $\md$. Halo mass is the primary driver of
$\pc_2$, though $\md$ contributes, now acting in opposition to
$\Mh$ rather than in concert with it.

\begin{figure}
\plotone{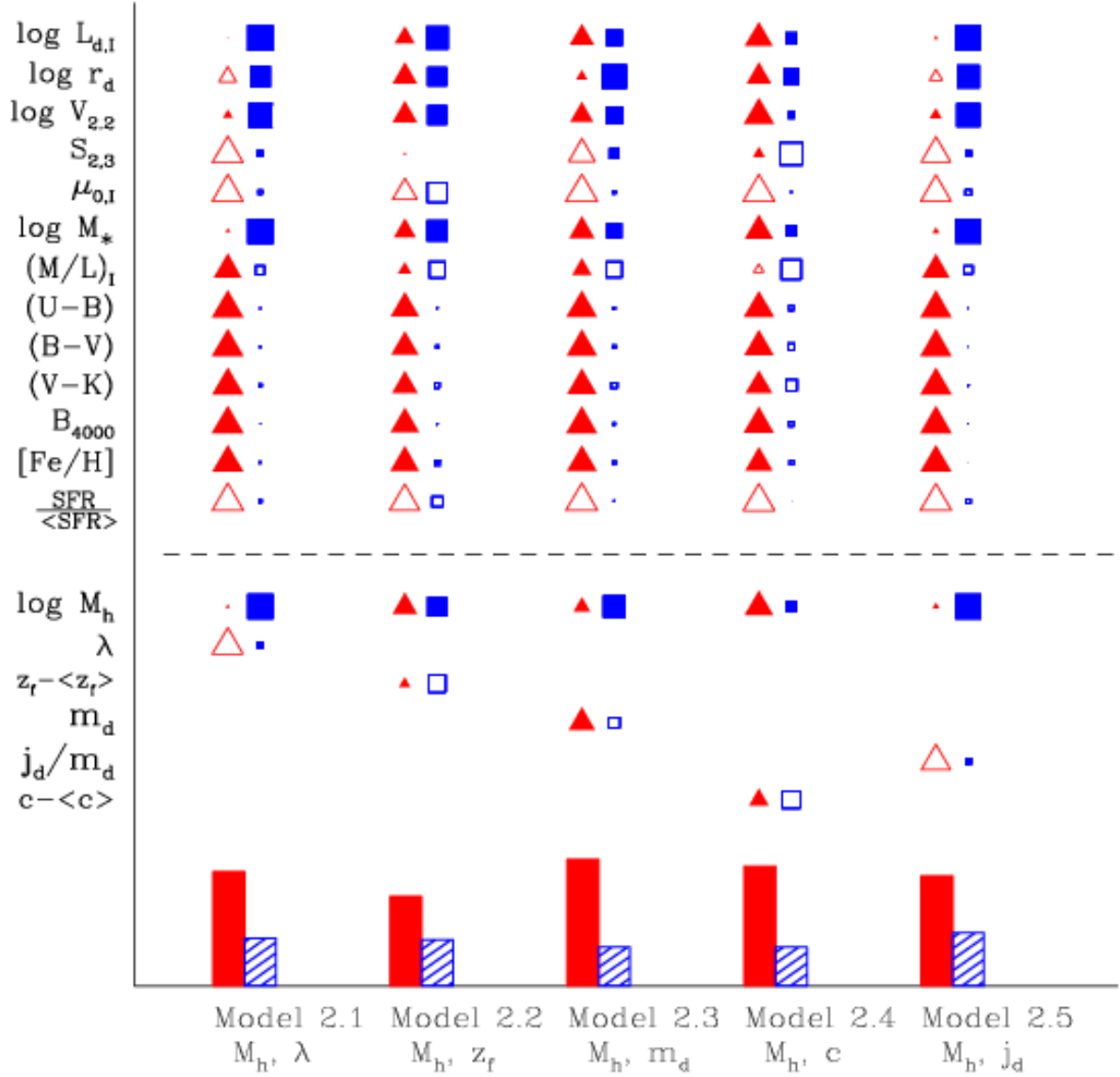}
\caption{
Summary of the PC structure of the five 2-parameter models.  For each
model, vertical bars at the bottom of the diagram indicate the fraction
of variance accounted for by the first two PCs.  In the upper part
of the diagram, triangles indicate the correlation of the 13 observables
with $\pc_1$; a filled symbol indicates positive correlation, an open
symbol anti-correlation, and the linear size of the symbol is proportional
to the value of the correlation coefficient.  Squares show the
correlation with $\pc_2$, in similar fashion.  Below the dashed line,
triangles and squares show correlations of the model control parameters
with the principal components.  In models with $z_f$ or $c$ as control
parameters, correlations are computed for 
$z_f-\langle z_f \rangle$ or $c-\langle c \rangle$, 
where $\langle z_f \rangle$ and $\langle c \rangle$ 
are the mean parameter values for the galaxies' halo mass $\Mh$.
\label{fig:xmassub}}
\end{figure}

We would like to be able to compare the PC structure of our models
directly to each other and to consider all of the observables
simultaneously, and this requires a more compact representation
than the one in Figures~\ref{fig:lcdm.scr.model2}-\ref{fig:lcdm.scr.model5}.
Figure~\ref{fig:xmassub} is our attempt to achieve such a representation.
It summarizes the results of the 2-parameter models we have examined
so far, and of two additional models, 2.4 and 2.5, in which halo 
concentration and disk angular momentum fraction are control parameters.
The bottom part of the
diagram depicts the strength of the significant PCs recovered. The height
of each box, solid for $\pc_1$ and shaded for $\pc_2$,
is directly proportional to the amount of variance
explained by each PC. 
The top part of the diagram, above the dashed line, displays the
correlations between all 13 observables and the PCs.
Triangles represent the correlations with $\pc_1$, and
the linear size of the triangle is directly proportional to 
the magnitude of the Spearman correlation coefficient;
filled triangles indicate positive
correlation, empty triangles negative correlation. 
Squares represent the correlations with $\pc_2$, in similar fashion.
Symbols below the dashed line display correlations of the model's 
control parameters with the PCs.  The mean values of
formation redshift $z_f$ and concentration parameter $c$ vary
with $\Mh$, so to isolate the impact of {\it variations} in these
quantities we subtract off the mean value $\langle z_f \rangle$
or $\langle c \rangle$ for the galaxy's halo mass before computing
correlation coefficients.  Similarly, the disk's specific angular
momentum, which determines its structural properties, is proportional
to $j_d/m_d$ rather than to $j_d$ itself, so we treat this ratio
as the control parameter.  As it happens, our models with varying 
$j_d/m_d$ have fixed $m_d$, and only rank values enter the
Spearman correlation coefficient, so our results would be no different
if we used $j_d$ instead, but the distinction would matter if 
we considered a model with independent variations in $j_d$ and $\md$.

The main shortcoming of Figure~\ref{fig:xmassub} is that it summarizes
correlations by a single number, the correlation coefficient.
As seen in Figures~\ref{fig:lcdm.scr.model2}-\ref{fig:lcdm.scr.model5},
moderate or weak correlations can have a variety of detailed structures ---
random scatter on top of a weak or non-existent trend, definition by
exclusion zones rather than a tight core, or washing out of a significant
underlying correlation by relatively rare outliers.
Figure~\ref{fig:xmassub} allows easy comparison of the overall 
PC structure of different models, at the price of losing this detailed
information for individual cases.

The first three columns of Figure~\ref{fig:xmassub} summarize what we have
seen in Figures~\ref{fig:lcdm.scr.model2}-\ref{fig:lcdm.scr.model5}. 
In each model, $\pc_1$ is largely a measure of SED shape; newly plotted
observables
$(U-B)$, $(V-K)$, and $B_{4000}$ are also strongly correlated with $\pc_1$.
The SED shape is always strongly correlated with surface brightness because
of the Schmidt-law connection between star formation rate and surface
density. (The sign of this correlation is negative because of the
mag arcsec$^{-2}$ units.)
$\pc_2$ is correlated mainly with scale variables like $\Ldi$,
$\Rd$ and $\vdmax$. However, the physical parameters determining $\pc_1$
are different in each case: $\lambda$ in Model 2.1, $M_h$ in Model 2.2,
and a combination of $m_d$ and $M_h$ in Model 2.3, with $m_d$ being
dominant.

Differences in the governing parameters lead to some important differences
in PCs. In Model 2.1, we see the dominance of the broad $\lambda$
distribution in determining the surface density. This leads to
anti-correlation of $\s23$ and $\Rd$ with $\pc_1$ and essentially no
correlation between $\Ldi$ and $\pc_1$.  
In Model 2.2, the greater importance of $M_h$ in
setting the surface density forces scale quantities to move partly into
$\pc_1$, and the sign of the correlation between $\Rd$ and $\pc_1$
reverses. In Model 2.3, $m_d$ variations link $\Ldi$ to 
surface density and thus to the shape of the SED. 
However, $\Rd$ remains in $\pc_2$ because it is determined mainly by $M_h$ 
(and by $\lambda$, which is fixed). $\vdmax$ has similar correlations
with $\pc_1$ in all three models, but for somewhat different reasons: 
in 2.1 and 2.3 because $\lambda$ or $m_d$ determines the surface density and
the baryon contribution to $\vdmax$, in 2.2 because $M_h$ determines
the surface density and the halo contribution to $\vdmax$.
The rotation curve slope is strongly anti-correlated with $\pc_1$
in the models where the baryon contribution to the rotation curve
varies substantially, but it is uncorrelated with either PC (and
hardly variable at all) in Model 2.2, where the baryon contribution
is constant.

The last two columns of Figure~\ref{fig:xmassub} show results for two more
2-parameter models. In Model 2.4, only $M_h$ and the concentration
parameter $c$ vary.  Most observables now have a stronger correlation
with $\pc_1$ than with $\pc_2$, the only exceptions being the 
rotation curve slope and the stellar mass-to-light ratio.
$\pc_1$ is driven primarily by halo mass, with a small contribution
from concentration.  $\pc_2$ depends on the opposite combination of
mass and concentration (anti-correlated instead of correlated), and
concentration dominates.  Less concentrated halos host galaxies with
flatter rotation curves, slightly larger disks, and (as a consequence
of slower star formation rates) lower $(M/L)_I$.

Model 2.5 has all parameters fixed except $M_h$ and $j_d$, the ratio of
disk to halo angular momentum.  In many ways, the physics of this model is
similar to that of Model 2.1, since only halo mass and disk angular
momentum vary.  The distribution of angular momentum is narrower in
this model --- in particular, there are no very large values because the
maximum angular momentum comes for $\lambda=0.05$, $j_d=1$.  
However, this difference in distributions does not have a large effect on 
the PC structure, and the correlations between the observables and
the PCs are nearly identical to those found for Model 2.1.

\subsection{The Baseline Model}
\label{subsec:minimalplus}

A realistic model of galaxy formation should include, at a minimum,
the expected variations in halo mass, spin parameter, and halo
formation redshift.  We define Model 3, with these three control
parameters, to be our ``baseline model.''
Figure~\ref{fig:matrix4fit}a shows bivariate correlations for this model.
The observed scatter is now determined by the variation
(at given mass) of $\lambda$ and $z_f$. 
The correlations and scatter in Figure~\ref{fig:matrix4fit}a are similar
to those of Model 2.1 shown in Figure~\ref{fig:matrix4subfit}b, 
indicating that $z_f$ variations (suppressed in Model 2.1) have
only small impact relative to $\Mh$ and $\lambda$.  
There are clear correlations among
$\Rd$, $\Ldi$ and $\vdmax$, with the scatter from $\lambda$ variations
especially evident in the R-V relation. Variations in formation redshift
add scatter to the correlations of color with other parameters,
but they do not erase them or change their sign, confirming the
dominant role of $\lambda$ in determining the star formation history
through the surface density.

\begin{figure}
\plotone{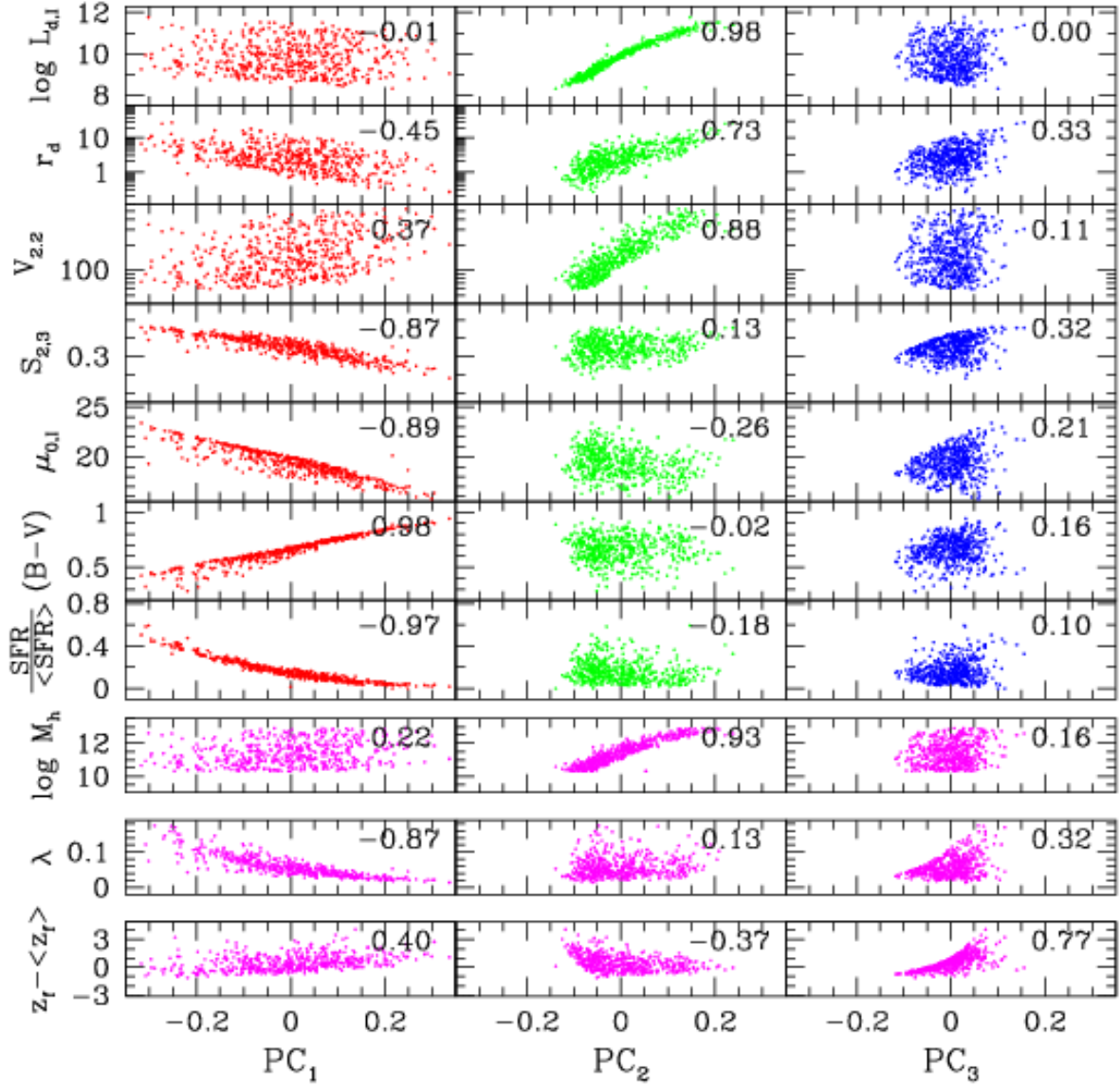}
\caption{
Correlations of observables (top) and control parameters (bottom)
with the three principal components of the baseline model, Model 3,
which has control parameters $\Mh$, $\lambda$, and $z_f$.
Format is the same as Fig.~\ref{fig:lcdm.scr.model2}.
\label{fig:lcdm.scr.model7}}
\end{figure}

\begin{figure}
\plotone{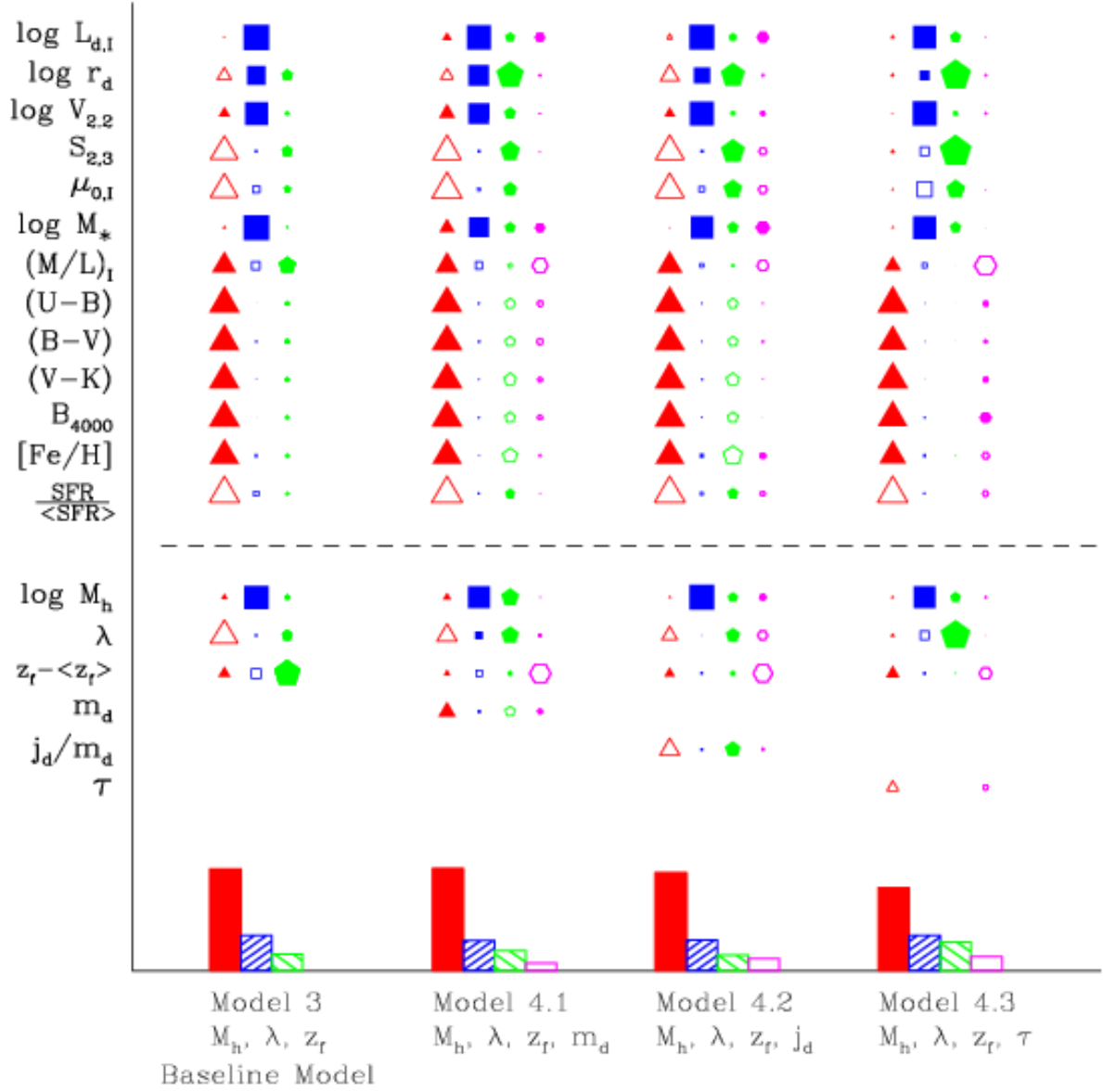}
\caption{
Principal component structure of the models with three or four
control parameters, in the same format as Fig.~\ref{fig:xmassub}.
Correlations with the $n$th PC are represented by symbols with $n+2$
sides, filled for positive correlation, open for anti-correlation,
with linear size proportional to the correlation coefficient.
The leftmost column encapsulates the predictions of the 
``standard'' theory of disk galaxy formation, our
baseline model, in which $M_h$, $\lambda$, and $z_f$ determine disk
structural parameters and the Schmidt law determines the star formation
history given these parameters. 
Other columns show results for models in which variations in
other physical processes contribute to variations in galaxy properties.
\label{fig:xmas}}
\end{figure}

PCs for the baseline model are illustrated in
Figure~\ref{fig:lcdm.scr.model7} and in the first column
of Figure~\ref{fig:xmas}.  Results are again reminiscent of those for
Model 2.1 (see Figure~\ref{fig:lcdm.scr.model2} and the first column of
Figure~\ref{fig:xmassub}), but the addition of $z_f$ as a control
parameter adds complexity.  The first PC is dominated as usual by SED 
quantities, which are strongly anti-correlated with $\mu_{0,I}$, weakly
correlated with $\vdmax$, moderately anti-correlated with $\Rd$,
and entirely uncorrelated with $\Ldi$.
The physical parameter driving $\pc_1$ is $\lambda$,
through its impact on surface density, though there are 
small contributions from $z_f$ and $\Mh$.
The greater importance of disk self-gravity and adiabatic contraction
in low-spin systems couples $\s23$ into $\pc_1$.

The second PC is again comprised mainly of scale quantities $\Ldi$, $\Rd$
and $\vdmax$. As in Model 2.1, $\pc_2$ is driven mainly by $M_h$, but 
now there is a slight contribution from $z_f$. 
The addition of a third control parameter leads
to the appearance of a third, relatively weak PC, involving 
the stellar mass-to-light ratio, the rotation curve slope, 
the disk scale length, and the central surface brightness.
$\pc_3$ is driven mainly by $z_f$ variations, 
and to successively smaller degrees by
$\lambda$ and $M_h$, but it does not have an obvious simple
interpretation. It seems to
arise mainly because $z_f$ and $\lambda$ compete in their contributions to
$\pc_1$, and a different combination of them (correlated instead of
anti-correlated) causes orthogonal variation in other observables. We
will discuss the PC structure of the baseline model further in
Section~\ref{subsec:summary}. 

Numerical and analytic studies of halo formation predict significant
variations in concentration, so one could argue that $c$ variations
should be included in our baseline model.  However, we have already
seen in Figure~\ref{fig:xmassub} that the influence of $c$ on PC
structure is relatively small.  We have investigated a case in
which we add $c$ variations to the baseline model, and the effects
are minor as expected.  The first two principal components are entirely
unchanged, but $\pc_3$ changes to some degree, splitting into two
weak PCs that involve similar observables in somewhat different combinations.
Halo concentration is not a major driver in any of these
principal components.  Because of its relatively small impact, we
have opted for simplicity and eliminated $c$ variations in our
baseline model and the extensions discussed below.

\subsection{Extended Models}
\label{subsec:extended}

\begin{figure}
\plotone{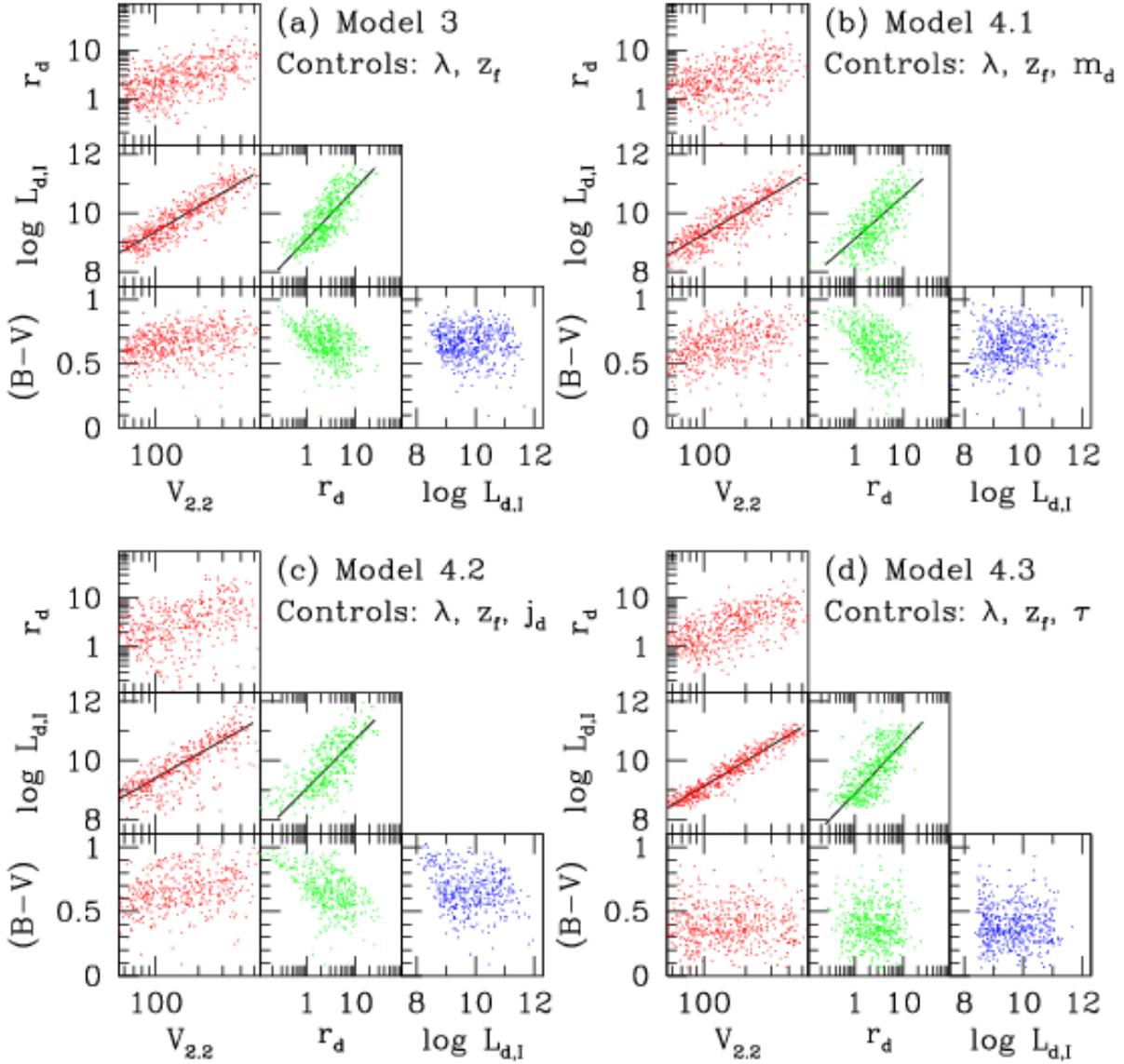}
\caption{
Bivariate correlations for models with three or four control
parameters, in the same format as Fig.~\ref{fig:matrix4subfit}.
Note that, in addition to the parameters listed, $\Mh$ is
a control parameter in every model.
Model 3 is the ``baseline model,'' with control parameters
$\Mh$, $\lambda$, and $z_f$.  
Models 4.1 and 4.2 extend the baseline model by adding $m_d$
or $j_d$ as an additional control parameter.  Model 4.3
decouples the star formation rate from the gas
surface density, adding a randomly chosen exponential decay
timescale $\tau$ as a control parameter.
\label{fig:matrix4fit}}   
\end{figure}

We can now ask what happens if we add new physical ingredients
to the baseline model, represented by additional control parameters.
Parameter correlations for three such extended models are illustrated in
Figures~\ref{fig:matrix4fit}b-d. Figure~\ref{fig:matrix4fit}b shows Model 4.1,
in which $m_d$ varies in addition to $M_h$, $\lambda$, and $z_f$.
This additional variation causes remarkably little change in the
correlations, relative to the baseline model shown in
Figure~\ref{fig:matrix4fit}a.  The scatterplots display a somewhat
increased number of outliers, but the cores of the correlations do not change.
This small impact of $m_d$ variations could be anticipated
on the basis of Figure~\ref{fig:matrix4subfit}d.

Figure~\ref{fig:matrix4fit}c shows a model with $m_d$ fixed at $0.5f_b$ 
but $j_d/m_d$ varying uniformly in the range $(0,1]$.   
Here the scatter in correlations is considerably larger than in the baseline
model because of the larger range in disk angular momentum.  
Nevertheless, the mean correlations are not fundamentally different from
those of Model 3.

Figure~\ref{fig:matrix4fit}d illustrates a more radical change to the
baseline model. Instead of the Schmidt law, we use an exponentially
decaying SFR beginning at $z_f$ with decay timescale 
$\tau$ chosen from a uniform
distribution $[1,9]$ Gyr, as described in \S\ref{subsec:sfr}.
This model specifically breaks the link
between the star formation history and the structural quantities that 
determine surface density.  As a result, color is
uncorrelated with $\vdmax$, $\Rd$ and $\Ldi$.

\begin{figure}
\plotone{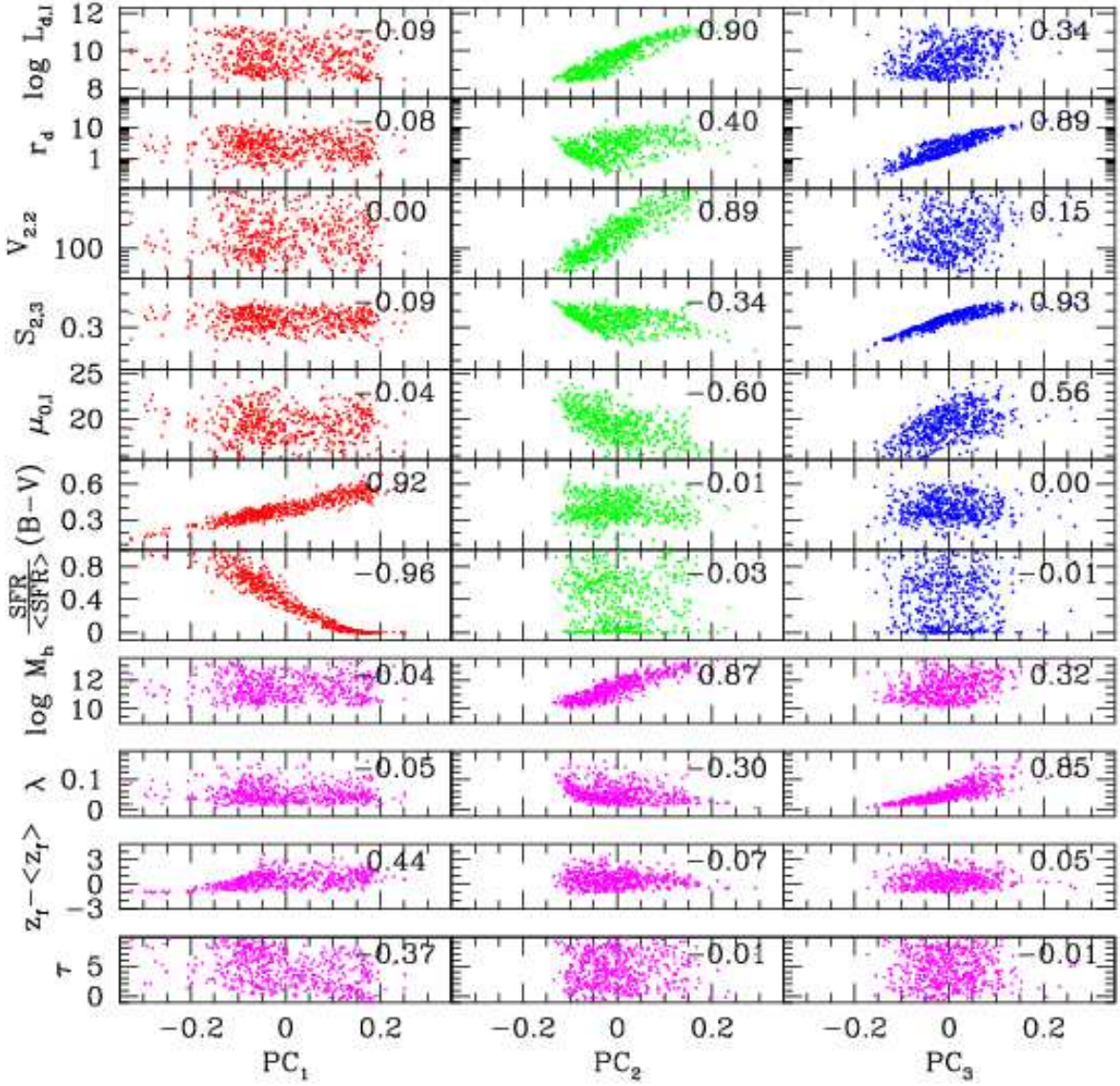}
\caption{
Same as Fig.~\ref{fig:lcdm.scr.model7}, but for Model 4.3, which decouples
star formation from surface density and adds the exponential star
formation timescale $\tau$ as a control parameter.
\label{fig:lcdm.scr.model16}}
\end{figure}

Figure~\ref{fig:lcdm.scr.model16} illustrates the PCs of this model.  
SED quantities still dominate the first component, and 
in contrast to the baseline model (Figure~\ref{fig:lcdm.scr.model7})
they are no longer correlated at all with $\mu_{0,I}$, $\Rd$, or $\vdmax$.  
Furthermore, $\pc_1$ is now determined by a combination of $z_f$
and $\tau$, rather than the combination of $z_f$ and $\lambda$ 
that drives $\pc_1$ in the baseline model. 
Also in contrast to the baseline model,
Model 4.3 has second and third PCs of nearly equal strength
(see the fourth column of Figure~\ref{fig:xmas}), both of
them composed mainly of scale quantities and surface brightness,
but in different combinations.  $\pc_2$ is largest when $\Mh$ is
large and $\lambda$ is small; this combination produces disks with
high luminosity, high circular velocity, and high surface brightness
(low $\mu_{0,I}$),
but the competing effects on scale length leave $r_d$ only 
moderately correlated with $\pc_2$.  $\pc_3$ is largest when $\Mh$
and $\lambda$ are {\it both} large; this combination produces disks
with large scale length and (less consistently) low surface
brightness and high luminosity.  However, the competing effects
on halo and disk contributions to the rotation curve leave $\vdmax$
almost uncorrelated with $\pc_3$.  Decoupling star formation from
surface density allows the correlated and anti-correlated combinations
of $\Mh$ and $\lambda$ to drive separate principal components of 
comparable strength, with $\Mh$ playing the stronger role in one ($\pc_2$)
and $\lambda$ in the other ($\pc_3$).
In the baseline model, by contrast, the strong
coupling of $\lambda$ into the SED principal component leaves mass as
the sole driver of $\pc_2$.  In terms of the observables, models 
with a Schmidt law prescription must have surface brightness 
strongly correlated with SED shape, and it is only abandoning this
prescription that allows correlated and anti-correlated trends of
surface brightness with other structural quantities to appear as
separate principal components.

\subsection{Summary}
\label{subsec:summary}
Figure~\ref{fig:xmas} summarizes our results for the baseline model and for the
extended models discussed in \S~\ref{subsec:extended}.  
The format is similar to that of Figure~\ref{fig:xmassub},
but since these models have three or four control parameters, there are
three or four significant PCs.

The leftmost column encapsulates the predictions of the ``standard'' theory of
disk galaxy formation (our baseline model), in which $M_h$, $\lambda$, and
$z_f$ determine disk structural parameters and the Schmidt law determines
the star formation history given these parameters. The first PC is
basically a measure of SED shape, strongly correlated with surface
brightness because of the Schmidt law. $\pc_1$ is also correlated with
$\vdmax$, and with $\s23$, because of the baryonic influence on
the rotation curve. The second PC consists mainly of scale quantities such
as $\Ldi$, $\Rd$, and $\vdmax$.  
Physically, the first PC is driven mainly by $\lambda$, the second mainly
by $M_h$.  Variations in $z_f$ drive the relatively weak third PC, 
which has correlated contributions from $\Rd$, $\s23$, $\mu_{0,I}$, 
and $(M/L)_I$.

One interesting if somewhat disappointing result of our analysis is that
extending the baseline model by adding stochastic variations in $\md$
makes little difference to the predicted PC structure (Model 4.1, 2nd column).
The first two PCs of Model 4.1 are nearly identical to those of Model 3,
though in this case $\pc_1$ is driven by a combination of $\md$ and
$\lambda$ instead of by $\lambda$ alone, and it picks up small 
contributions from $\Ldi$ and $M_\star$ as a result.  The third PC
is noticeably different, having stronger correlations (especially
with $\Rd$ and $\s23$) and a non-negligible contribution from SED
observables.  The addition of $\md$ as a variable that influences
surface density allows a correlated contribution of $\Mh$ and $\lambda$
to drive larger variations in $\Rd$, which are correlated in turn
with $\s23$ because extended disks have less dynamical impact.
The fourth PC, driven by $z_f$ variations, is very weak; this is
the only case where we plot a PC with variance less than
our statistically motivated threshold of 0.7 (see \S\ref{sec:pca}).

As discussed in \S\ref{subsec:subminimal} with regard to Model 2.3,
$\md$ variations have less impact on the scatter of the L-V relation
than one might expect because disk gravity effects tend to shift points
parallel to the L-V locus as $\md$ changes.  
The mean value of $\md$
should have a noticeable effect on the zero points of some relations,
such as L-R, and on the relative velocity scales inferred at fixed
luminosity from rotation curves (which probe the regime where disk
gravity is important) and from weak lensing measurements (which probe
larger scales).  However, the effect of scatter in $\md$ is harder
to discern.  
The differences in $\pc_3$ between Models 3 and 4.1 are
probably large enough to be observable, but it is less clear that
they are larger than uncertainties associated with the approximations
of our modeling.  At the least, comparison of Models 3 and 4.1 implies
that agreement between observed correlations of galaxy properties and 
predictions of the baseline model, if found in the data, should not be 
taken as immediate evidence that $m_d$ is constant from halo to halo.
Measurements of rotation velocity at somewhat larger radii, $3-5\Rd$, could
be helpful in revealing effects of $\md$ scatter, since they are less
influenced by the gravity of the disk \citep{she2001}.

Adding variations in angular momentum loss to the baseline
model also has little effect on the first two principal components
(Model 4.2, 3rd column).  It is now a combination of $\lambda$
and $j_d$ that drives $\pc_1$, and $\Mh$ still drives $\pc_2$.
The broader angular momentum
distribution increases the scatter in bivariate relations
(see Figure~\ref{fig:matrix4fit}), but the overall correlation
structure is much the same.  The third PC, on the other hand, resembles
that of Model 4.1 rather than Model 3, and it is driven again by
a correlated contribution of halo mass and disk angular momentum.
Formation redshift provides the main contribution to the fourth PC,
which is similar to that of Model 4.1.
However, $\pc_4$ has no strong correlation with any of the observables,
and its statistical significance is marginal.

Breaking the Schmidt-law link between star formation and surface density
makes an important difference to the PC structure, leading to a clean
separation between SED quantities and structural quantities (Model 4.3,
4th column).  Central surface brightness and rotation curve slope
vanish from $\pc_1$, where they had a strong presence in all of
our previous models.  The weak contributions of $\Rd$ and $\vdmax$
also disappear.  The coherent SED variations that comprise $\pc_1$
are driven mainly by formation redshift and the
exponential decay timescale.  There are now two PCs comprised
of scale quantities, both of nearly equal strength, driven by $M_h$ and
$\lambda$ in two different combinations (correlated and anti-correlated).
The first of these contains correlated contributions of $\Ldi$ and
$\vdmax$ and the anti-correlated contribution of $\mu_{0,I}$, but it has
only a moderate contribution from $\Rd$.
The second represents correlation of $\Rd$, $\s23$ and $\mu_{0,I}$, 
with moderate contribution from $\Ldi$.  Finally, $PC_4$,
associated with $z_f$ and $\tau$, accounts for some of the
variance in $(M/L)_I$. 

\section{Conclusions and Outlook}
\label{sec:the_end}

The techniques describe in \S\ref{sec:overview} and \S\ref{sec:pca}
allow us to predict the principal component structure of the disk galaxy 
population, given different assumptions about the control parameters
that govern the origin of galaxies' observable properties
(see Tables~\ref{tab:input} and~\ref{tab:models}).
By examining the correlation of the principal components with the input
parameters, we also learn what physical processes drive these 
components in different galaxy formation models.  Our list
of observables (Table~\ref{tab:output}) includes three broad
band colors, the 4000\AA\ break, and the birth parameter $b=\birth$,
and these quantities are highly correlated with each other because
they are all determined by the galaxy's star formation 
history.\footnote{\cite{con1995} show that the optical spectra of
galaxies form something close to a 1-parameter family, so this
high degree of correlation is a property of observed galaxies and
not simply an artifact of our modeling procedures.}  
As a result, these SED parameters dominate the first principal
component, $\pc_1$, in all of our models.  $\pc_2$ is usually a 
measure of overall scale, with strong contributions from luminosity,
circular velocity, disk scale length, and stellar mass.

Distinctions among our models appear in the apportioning of these
scale observables and two other structural quantities,
the central surface brightness and rotation curve slope,
among $\pc_1$ and $\pc_2$, and in some cases $\pc_3$.
These apportionments depend in turn on the governing physics
of the model.  In all models that have a Schmidt-law prescription
for star formation, the central surface brightness is strongly
correlated with $\pc_1$ because of the strong coupling between
star formation history and surface density.
In models where angular momentum and/or disk mass fraction control
the surface density, high surface brightness disks also have
large baryon contributions to the rotation curve, producing a strong
coupling to $\s23$.  Our baseline model has $\Mh$, $\lambda$, and
$z_f$ as control parameters, and because the log-normal distribution
of $\lambda$ is very broad, it dominates the distribution of disk
surface densities and drives $\pc_1$.  Halo mass drives $\pc_2$,
and formation redshift drives a weak third component.  When
$\md$ or $j_d$ become additional control parameters, they share
direction of $\pc_1$ with $\lambda$, and this allows a correlated
combination of halo mass and angular momentum to produce a new third PC that
has substantial correlations with $\Rd$ and $\s23$.  The most
significant change to the PC structure comes from replacing the
Schmidt law by an exponentially declining star formation prescription
with decay timescale chosen independently of surface density.
Surface brightness and $\s23$ disappear from the first principal
component, and correlated and anti-correlated combinations of
$\Mh$ and $\lambda$ drive two different PCs of nearly equal
strength, each involving a combination of structural quantities.
Thus, PCA of the observed disk galaxy population can distinguish
among different models for the origin of galaxy properties.

There are numerous ways to improve or extend our models of disk
galaxy formation and thus provide a more comprehensive framework
for understanding the implications of PCA.  One of the most
obvious ingredients missing from our calculations is a model
of dust extinction and scattering, which can have important 
effects on luminosities, colors, and mass-to-light ratios.
Since the impact of dust is highly dependent on inclination,
one would need to include inclination or axis ratio as an additional
observable in models that incorporate dust.  Alternatively,
one can determine empirical corrections for internal extinction
and correct the data to values for face-on disks.  Given the complexities
and uncertainties of realistic dust modeling (see, e.g.,
\citealt{woo1997,sil2001}), this empirical correction approach,
already commonly used in Tully-Fisher analyses (see \citealt{tul1985}),
may be preferable to adding dust to the models.

Since the star formation prescription plays such a fundamental
role in governing the PC structure of our models, it would be
interesting to explore alternatives to the Schmidt law that are
not as extreme as our exponential decay model, which decouples
star formation from structural quantities completely.
The assumption that star formation continues until the 
effective local value of the \cite{too1964} instability parameter
$Q$ exceeds some threshold (see \citealt{gun1981,gun1983,gun1987}) 
is one example of such an 
alternative, and the role of disk self-gravity in determining $Q$
might lead to a significantly different correlation between SED
and structural quantities in $\pc_1$.
Our models also subsume
all of the complex physics of gas cooling and feedback into
the single parameter $\md$.  If these processes are sufficiently
self-regulating that $\md$ is nearly constant, or if variations
of $\md$ are truly stochastic, then our simple description may well
be adequate to our purpose.  However, one could imagine that a 
more detailed model of cooling and feedback would connect variations
in disk mass fraction to variations in star formation history or galaxy 
structure, and that these connections might in turn alter the structure
of principal components.

Fundamental to our models is the assumption that disk sizes are
determined by the combination of halo virial radii with angular
momentum parameters ($\lambda$ and in some cases $j_d$) that vary
independently of other model inputs.  The influence of angular
momentum on the galaxy's stellar population is ``one way,'' with
disk scale length determining the star formation rate through
the Schmidt law.  This assumption seems fairly plausible even if the
process of disk assembly is messier than the one envisioned in
our adopted formalism, but it is not 
incontrovertible.  While the theoretically predicted distribution
of $\lambda$ yields rough agreement with the observed distribution
of disk sizes \citep{sye1999,dej2000}, the detailed distribution of
specific angular momentum within halos would not, if preserved by
the collapsing baryons, yield exponential disks \citep{ryd1988,bul2001a,van2001}.
An alternative explanation for exponential stellar disks is viscous
redistribution of disk material on the same timescale as star
formation \citep{lin1987,sly2001}.  A model incorporating this kind 
of ``back reaction'' of star formation on scale length might make 
distinctive predictions for PC structure.  The dynamical interactions
between baryons and the dark halo could well be more complicated
than the adiabatic contraction model we have used, for example
because of resonant interactions with rotating bars 
\citep{her1992,wei2001},
and such interactions might also alter PCA predictions in distinctive ways.

The most important and challenging direction for extending our
models is to incorporate bulge formation and transformation among
morphological types.  In contrast to disk formation, there is 
no ``standard model'' of bulge formation, though the idea that mergers
transform stellar disks into spheroids is the one that has
been most widely explored in semi-analytic models 
\citep{bau96b,kau1996,som1999}.
Alternative ideas include secular bulge formation via bar
instability \citep{rah1991,van2001}, association of rapid early
star formation with spheroids and slower subsequent star 
formation with disks \citep{egg1962}, and morphological
transformation in groups and clusters
caused by weak dynamical perturbations \citep{moo1998}
or interactions with intergalactic gas \citep{gun1972}.
Galaxy formation models that incorporate
these processes could predict the PC structure of the full galaxy
population instead of the isolated disk systems that we have
considered, and they would show whether PCA can diagnose the
relative importance of different mechanisms for morphological
transformation.  The calculational approach needed for such models
is more complicated than the one we have employed in this paper,
requiring halo and galaxy merger trees.  One fortunate by-product
of such an approach would be descriptions of the local environment ---
field, group, cluster --- of each model galaxy.  
These environmental descriptions could be incorporated as additional
observables in PCA, and they would likely add considerable power
for diagnosing the origin of morphological properties.

The advantages of semi-analytic models for PCA predictions are
computational speed, the relatively transparent links between
input parameters and output observables, and the ease with
which one can vary model assumptions and examine their impact
on PC structure.  However, hydrodynamic simulations incorporate
more realistic descriptions of some of the essential physical
processes --- gravitational collapse, gas dynamics and cooling,
galaxy mergers within common halos --- and they are approaching
the point where they could usefully be analyzed with the same 
multi-variate techniques employed here.  High resolution 
simulations of individual galaxies (e.g., \citealt{nav2000})
can predict many of the quantities that we have used in our
analysis, and improvements in computer hardware and algorithms
should eventually allow creation of the kinds of ensembles that
would be needed for PCA.  An intermediate option is to use somewhat
lower resolution simulations of large volumes 
(e.g., \citealt{pea1999,mur2001,nag2001,yos2001}) to predict
the baryonic masses, assembly histories, environments, and
total angular momenta of galaxies, and to supplement these with
models of the sub-resolution physics that translates these
global quantities into direct observables.  The combination
of numerical and semi-analytic approaches could be fruitful,
with numerical simulations calibrating the 
semi-analytic calculations for matched assumptions and 
semi-analytic calculations illustrating how changes to the 
input assumptions might alter the numerical predictions.

The recent convergence on a well defined cosmological model
has solidified the ground beneath a basic picture of galaxy
formation that has emerged over the last 25 years: galaxies
form after gas collapses and dissipates within collisionless
dark matter halos, which form by gravitational
instability from initial conditions that are
not far from those of the $\Lambda$CDM scenario.
However, there are still many competing ideas for the origin
of galaxy luminosities, sizes, colors, morphologies, and dynamical
properties, and even the leading ideas have not been tested
extensively against observations.  As illustrated in this paper,
any theory that predicts these observable quantities also 
predicts the correlations among them, which can be well summarized
using the techniques of Principal Component Analysis.
Comparison of these predictions to PCA results from the SDSS
and other large surveys should take us a long way towards understanding
how the observable properties of galaxies are connected to the 
governing physics of galaxy formation.

\acknowledgments
We wish to thank Stephane Charlot, Laura Padovani and Paola Marigo for
providing their data in electronic form; Andreas Berlind, James Bullock
and Andrew Connolly for insightful and stimulating comments
that improved the overall presentation of this work;
and James Gunn and Robert Lupton for many useful discussions
of the SDSS and the use of PCA as a tool for characterizing the
galaxy population. AC has been supported by NASA through grant number
AR-07535.01-96A from the Space Telescope Science Institute, which is
operated by Association of Universities for Research in Astronomy,
Incorporated, under NASA contract NAS5-26555. AC was also supported by
NSF grant number AST-9519324.  DW acknowledges the hospitality of the
Institute for Advanced Study and financial support of the Ambrose Monell
Foundation during the final stages of this work.

\appendix
\section*{Appendix}
\section{Chemical Enrichment}
We compute the chemical enrichment in each model galaxy by solving the
following integro-differential equation:
\beq
\label{eq:feh_evol}
M_g \frac{dZ}{dt} = E_{Z_{new}}(t) + E_{Z_{old}}(t) - 
Z(t) [E(t) + F(t)]\ ,
\eeq
\noindent
where $M_g$ is the available gas content, $Z(t)$ is the metal abundance by
mass, $F$ is the infall rate of pristine gas, and $E$ is the ejection
rate of matter by evolved stars. 
The choice of an infall rate is more problematic than the choice of SFR,
since both theoretical and observational constraints are much less
obvious. Here we assume that pristine gas falls onto the disk at a rate
exponentially decreasing with time. Stars are formed at a rate
proportional to the surface gas density \citep{lac1983}. 
In order to reproduce the observed abundance gradients, \citet{lac1983}
suggested that the infall timescale should be an increasing function of
galactocentric radius. We assume an exponential form of the infall time
scale dependence $\tau(r) = \tau_{0} \exp(r/\Rd)$, where $\Rd$ is the disk
scale length. The characteristic timescale $\tau_0$ depends on
the total mass of the disk through the relation $\tau_{0} =
\tau_{\odot}(M_d/M_{d,MW})^{-1/2}$, where the subscript MW refers to the
Milky Way and $\tau_{\odot} = 4\;$Gyr is the collapse timescale for the
solar neighborhood \citep{mol1999,fer1994}. We further assume the infalling
gas to be at primordial metallicity.

The term representing the gas ejection rate in equation
(\ref{eq:feh_evol}) can be defined as a simple function of SFR, IMF,
metallicity, and mass: 
\beq
\label{eq:E}
E(t) = \int^{m_u}_{m_{(m,\zstar)}} 
m p(m,\zstar) \phi(m)
\psi(t-\tau(m,\zstar)) dm \ .
\eeq
\noindent
Here $p(m,\zstar) = [m - m_{\rm rem}(m,\zstar)]/m$ is the returned
mass fraction of stars of mass $m$ and remnant mass $m_{\rm rem}$,
$\psi(t-\tau(m,\zstar))$ is the SFR by number at the time a star of
initial mass $m$ and metallicity $\zstar$ was formed, $\tau(m,\zstar)$
is the lifetime of a star of initial mass $m$ born with metallicity
$\zstar$, and $m_t(m,\zstar)$ is the current turnoff mass.

The total ejection rate of metals can be expressed as the sum of two
terms: the first term takes into account the newly synthesized and
ejected metals, the second accounts for the rate of ejection of
unprocessed metals, i.e. those that originate from the material out of
which the star was formed. In detail:
\bea
\label{eq:Ez}
E_{Z_{new}}(t) & = & \int^{m_u}_{m_{(t,\zstar)}} m p_z(m,\zstar)
\phi(m) \psi(t-\tau(m,\zstar)) dm \\
E_{Z_{old}}(t) & = & \int^{m_u}_{m_{(t,\zstar)}} m p(m,\zstar)
Z(t-\tau(m,\zstar)) \phi(m) \psi(t-\tau(m,\zstar)) dm \ , 
\eea
\noindent
where $p_z(m,\zstar)$ is the heavy element integrated stellar yield,
which is defined relative to the initial metal abundance of the star, and
$\zstar = Z(t-\tau(m,\zstar))$ is the initial metal abundance of
stars that evolve off the red giant branch at age $t$. From
equations (\ref{eq:feh_evol}), (\ref{eq:E}), and (\ref{eq:Ez}), it is
evident that the present metal abundance is very sensitive
to the solution of the metallicity equation in the past. 

Direct computation of $E(t)$ using equation (\ref{eq:E})
cannot be achieved without detailed knowledge of $Z(t)$, unless
a numerical (iterative) procedure is used. Recently, however,
\citet{vdH1997} showed that if one assumes that at any time
during the galaxy evolution $Z(t-\tau(m,\zstar)) \leq Z(t)$, then
$E_{Z_{old}}(t) \leq Z(t)E(t)$. If we now define $E_{Z_{old}}(t) -
Z(t)E(t) \equiv G(t)Z(t)$, equation~(\ref{eq:feh_evol}), which describes
the chemical enrichment, can be written as \citep{vdH1997}:
\bea
\label{eq:feh_evol_new}
\frac{dZ}{dt} & = & \frac{1}{M_g(t)} \left\{ -Z(t) [G(t)+F(t)] +
E_{Z_{new}}(t) \right\} \\
              & = & -Z(t)P(t) + Q(t) ~,
\eea
\noindent
where
\bea
\label{eq:PandQ}
P(t) & = & \frac{1}{M_g(t)} \left[ G(t) + F(t) \right]\\
Q(t) & = & \frac{1}{M_g(t)} E_{Z_{new}}(t) ~.
\eea
\noindent
Finally, by integrating over time, we obtain the general solution for
the gas metallicity at time $t$:
\beq
\label{eq:Z}
Z(t) = e^{-\int^t P(\tau) \, d\tau} \times \left[Z(t=0) + 
\int^t \frac{Q(\tau)}{e^{-\int^t P(\tau') \, d\tau'}} \, d\tau \right] ~.
\eeq 
\noindent
All integrals are computed
taking explicitly into account the finite
lifetime $\tau_m$ of a star of mass $m$ and the metal abundance
$\zstar$ at its formation time $(t-\tau_m)$. We used metallicity
dependent stellar lifetimes and yields from the chemically consistent
models of \citet{mar1998}, for stellar masses $m<6 \msun$, and of
\citet{por1998}, for stellar masses up to $125 \msun$. Both these authors
give tables of lifetimes and yields as a function of mass for five
discrete metallicities:
$Z=0.0004$, $Z=0.004$, $Z=0.008$, $Z=0.02$, $Z=0.05$. We then 
linearly interpolate in $m$ and $Z$ where necessary. We further assume
that each star expels its ejecta all at once at the end
of its lifetime, and that the ejected material is immediately mixed in the
ISM, which remains always homogeneous. It is worth mentioning that this
``instantaneous mixing approximation'' is only suitable to reproduce
average trends observed in the age--metallicity relation and abundance
ratios \citep{por1998,vdH1997a}.

\clearpage
{}

\clearpage
\begin{deluxetable}{l}
\tablecolumns{1}
\tablewidth{0pc}
\footnotesize
\tablecaption{Theoretical parameters that define a galaxy.\label{tab:input}}
\tablehead{\colhead{Inputs}}
\startdata
Halo Mass: $M_h$ \\

Halo Concentration Parameter: $c$\\

Halo Spin Parameter: $\lambda$ \\

Halo Formation Redshift: $z_f$ \\

Disk Mass Fraction: $\md = M_{d}/M_{h}$ \\

Disk Angular Momentum Fraction: $\jd = J_{d}/J_{h}$ \\

E-folding time of exponentially decaying SFR$^{\dag}$: $\tau$\\
\enddata
\tablenotetext{\dag}{Only in those realizations where the SFR is {\it
not} set by the empirical Schmidt law.}
\end{deluxetable}

\begin{deluxetable}{l}
\tablecolumns{1}
\tablewidth{0pc}
\footnotesize
\tablecaption{Observable Outputs$^{\dag}$\label{tab:output}}
\tablehead{\colhead{Outputs}}
\startdata
I-band Luminosity (log): $\log \Ldi$ $[L_\odot]$\\
Disk Scale Length (log): $\log \Rd$ [kpc] \\
Rotational Velocity at Disk's Maximum (log): $\log \vdmax$ $[\kms]$\\
Rotation Curve Slope: $\s23$ \\
I-band Central Surface Brightness: $\mu_{0,I}$ [mag arcsec$^{-2}$]\\
Stellar Mass (log): $\log M_{\star}$ $[M_\odot]$ \\
I-band Stellar Mass-to-Light Ratio: $M_{\star}/L_I$ $[M_\odot/L_\odot]$\\
Broad-band Colors: $(U-B)$, $(B-V)$, $(V-K)$ \\
Amplitude of 4000\AA \ break: $B_{4000}$ \\
Mean Metallicity: $[{\rm Fe/H}]$ \\
Birth Parameter: $b$ = $\birth$ \\
\enddata
\tablenotetext{\dag}{List of observable quantities produced for each
galaxy as a function of the theoretical parameters in
Table~\ref{tab:input}. Each quantity is followed as a function of
redshift throughout the lifetime of the galaxy. These observables are
then used as inputs for the Principal Component Analysis.
Units are listed for dimensional quantities.  A (log) notation 
indicates that the logarithm of the observable is used in the PCA
rather than the linear variable.}
\end{deluxetable}

\begin{deluxetable}{l|c|c|c|c|c|c|c}
\tablewidth{0cm}
\tablecaption{Model Parameters$^{\dag}$\label{tab:models}}
\tablehead{
\colhead{Control Parameters} & \colhead{$z_f$} & 
\colhead{$m_d/f_b$} & \colhead{$j_d/m_d$} & 
\colhead{$\lambda$} & \colhead{c} & \colhead{SFR}}
\startdata
Model 1: $\Mh$                  
  & $\bar{z}_f(M_h)$  & 0.5 & 1 & 0.05 & ${\bar c}(\Mh)$ & Kennicutt\\
Model 2.1: $\Mh$, $\lambda$ 
  & $\bar{z}_f(M_h)$  & 0.5 & 1 &  -1  & ${\bar c}(\Mh)$ & Kennicutt \\
Model 2.2: $\Mh$, $z_f$  
  & -1   & 0.5 & 1 & 0.05 & ${\bar c}(\Mh)$ & Kennicutt \\
Model 2.3: $\Mh$, $\md$   
  & $\bar{z}_f(M_h)$  & -1   & 1  & 0.05 & ${\bar c}(\Mh)$ & Kennicutt \\
Model 2.4: $\Mh$, $c$  
  & $\bar{z}_f(M_h)$  & 0.5 & 1 & 0.05 & -1 & Kennicutt \\
Model 2.5: $\Mh$, $\jd$     
  & $\bar{z}_f(M_h)$  & 0.5 & -1  & 0.05 & ${\bar c}(\Mh)$ & Kennicutt \\
Model 3: $\Mh$, $\lambda$, $z_f$     
  & -1 & 0.5 & 1  & -1 & ${\bar c}(\Mh)$ & Kennicutt\\
Model 4.1: $\Mh$, $\lambda$, $z_f$, $\md$ 
  & -1 & -1 & 1 & -1 & ${\bar c}(\Mh)$ & Kennicutt\\
Model 4.2: $\Mh$, $\lambda$, $z_f$, $\jd$ 
  & -1 & 0.5 & -1  & -1 & ${\bar c}(\Mh)$ & Kennicutt\\
Model 4.3: $\Mh$, $\lambda$, $z_f$, $\tau$ 
  & -1 & 0.5 & 1  & -1 & ${\bar c}(\Mh)$ & Exponential\\
\enddata
\tablenotetext{\dag}{
Parameter values for the ten galaxy formation models studied in 
\S\ref{sec:results}.  Entries of $-1$ indicate that the parameter values 
were chosen from a distribution: the LC predicted distribution for $z_f$,
uniform distributions in the interval $(0,1]$ for $m_d/f_b$ or $j_d/m_d$,
and log-normal distributions with the numerically predicted means
and variances for $\lambda$ and $c$.  In every model, halo masses are 
drawn from a PS mass distribution truncated at circular velocities $\v200$
of $40\;\kms$ and $300\;\kms$.  Notations $\bar{z}_f(M_h)$ and 
$\bar{c}(M_h)$ indicate that $z_f$ and $c$ were fixed to the characteristic
value for the galaxy's halo mass.}
\end{deluxetable}



\clearpage

\clearpage

\clearpage

\clearpage

\clearpage

\clearpage

\clearpage

\clearpage

\clearpage

\clearpage


\end{document}